\documentclass[letterpaper,floatfix,superscriptaddress,preprint]{revtex4}
\usepackage{latexsym}
\usepackage{amstext,amsfonts}
\usepackage{epsfig}
\usepackage{natbib}
\usepackage{amssymb}
\usepackage{latexsym}
\usepackage{amsmath}
\begin{document}

\title{The Size Variance Relationship of Business Firm Growth Rates}

\author{Massimo Riccaboni}
\affiliation{DISA, University of Trento, Via Inama 5, Trento, 38100, Italy}
\affiliation{Center for Polymer Studies, Boston University, Boston,
Massachusetts 02215, USA}
\author{Fabio Pammolli}
\affiliation{IMT Institute for Advanced Studies, Via S. Micheletto 3, Lucca, 55100}
\affiliation{Center for Polymer Studies, Boston University, Boston,
Massachusetts 02215, USA}
\author{Sergey~V.~Buldyrev}
\affiliation{Department of Physics,~Yeshiva University, 500 West 185th Street,~New York, NY 10033 USA}
\author{Linda Ponta}
\affiliation{DIFIS, Politecnico di Torino, Corso Duca degli Abruzzi 24, Torino, 10129 Italy}
\author{H. Eugene Stanley}
\affiliation{Center for Polymer Studies, Boston University, Boston,
Massachusetts 02215, USA}

%\date{5 April 2009 --- pdcsh.tex}

\date{\today}% It is always \today, today,

\begin{abstract}
  The relationship between the size and the variance of firm growth rates is known to follow an approximate power-law behavior $\sigma(S) \sim S^{-\beta(S)}$ where $S$ is the firm size and $\beta(S)\approx 0.2$ is an exponent weakly dependent on $S$. Here we show how a model of proportional growth which treats firms as classes composed of various number of units of variable size, can explain this size-variance dependence.  In general, the model predicts that $\beta(S)$ must exhibit a crossover from $\beta(0)=0$ to $\beta(\infty)=1/2$. For a realistic set of parameters, $\beta(S)$ is approximately constant and can vary in the range from 0.14 to 0.2 depending on the average number of units in the firm. We test the model with a unique industry specific database in which firm sales are given in terms of the sum of the sales of all their products.
  We find that the model is consistent with the empirically observed size-variance relationship.
\end{abstract}

\pacs{89.75.Fb | 05.70.Ln  | 89.75.Da  | 89.65.Gh}

%\end{frontmatter}

\maketitle

\section{Introduction}
\label{sec:Introduction}

%\chapter{The Relationship between Size and Growth}~\label{Scaling}

Gibrat was probably the first who noticed the skew
size distributions of economic systems \cite{Gibrat31}. As a simple candidate explanation he postulated
the \textquotedblleft Law of Proportionate Effect\textquotedblright according to which
the expected value of the growth rate of a business firm is
proportional to the current size of the firm \cite{Sutton97}.
Several models of proportional growth have been subsequently
introduced in economics \cite{Kalecki45,Steindl65,Simon77,Sutton98}. In particular, Simon and
collegues \cite{Simon55,Simon75} examined a stochastic process for
Bose-Einstein statistics similar to the one
originally proposed by Yule \cite{Yule25} to explain the
distribution of sizes of genera. The Simon model is a Polya Urn
model in which the Gibrat's Law is modified by incorporating an entry process of new firms.
In Simon's framework, the firms capture a sequence of many independent
``opportunities'' which arise over time, each of size unity, with a
constant probability $b$ that a new opportunity is assigned to a new
firm. Despite the Simon model builds upon the Gibrat's Law, it leads to a skew
distribution of the Yule type for the
upper tail of the size distribution of firms while the limiting distribution
of the Gibrat growth process is lognormal.
The Law of Proportionate Effect implies that the variance $\sigma^2$
of firm growth rates is independent of size, while according to the Simon model it is
inversely proportional to the size of business firms. 
Hymer, Pashigian and Mansfield \cite{Hymer62,Mansfield62} noticed that the
relationship between the variance of growth rate and the size of
business firms is not null but decreases with increase in size of firm
by a factor less than $1/K$ we would expect if firms
were a collection of $K$ independent subunits of approximatively equal size.
In a lively debate in the mid-Sixties Simon and Mansfield \cite{SimonD64} argued that this
was probably due to common managerial influences and other similarities of firm units
which implies the growth rate of such components to be
positive correlated. On the contrary, Hymer and Pashigian \cite{HymerD64} maintained that larger firms
are riskier than expected because of economies of scale and monopolistic power.
Following Stanley and colleagues \cite{Stanley96} several 
scholars \cite{Bottazzi01,Sutton02} have recently found a non-trivial relationship 
between the size of the firm $S$ and the variance $\sigma^2$ of its growth 
rate: 

\begin{equation}
\sigma \sim S^{-\beta} \label{e:sigma}
\end{equation}
with $\beta \approx 0.2$.

Numerous attempts have been made to explain this puzzling evidence
%Eq. (\ref{e:sigma})
%Since it has been found that the correlation among the firm
%constituent parts is too weak to account for Eq. (\ref{e:sigma}) ~\citep{Bottazzi01,Sutton02},
%most of the researchers in this field have
by considering firms as collection of independent units
of uneven size \cite{Stanley96,Sutton02,DeFabritiis03,Sergey_II,Amaral97,Aoki07,Axtell06,Klepper06} 
but existing models do not provide a unifying explanation for the probability density functions of the
growth and size of firms as well as the size variance relationship.

Thus, the scaling of the variance of firm growth rates has been considered to be a crucial unsolved problem in economics \cite{Gabaix99,Sutton07}. Recent papers \cite{Fu_PNAS,Growiec07,Growiec08,Buldyrev07,Pammolli07} provide a general framework for the growth and size of business firms based on the number and size distribution of their constituent parts \cite{Sutton02,DeFabritiis03,Sergey_II,Amaral97,Amaral98,Takayasu98,Canning98,Buldyrev03}.
Specifically, Fu and colleagues \cite{Fu_PNAS} present a model of proportional growth in both the number of units and their size, drawing some general implications on the mechanisms which sustain business firm growth.
According to our model, the probability density function (PDF) of the growth rates
is Laplace in the center with power law tails. The PDF of the firm growth rates is markedly different from a lognormal
distribution of the growth rates predicted by Gibrat and comes from a convolution
of the distribution of the growth rates of constituent units and the distribution of number of units in economic systems.
The model by Fu and colleagues \cite{Fu_PNAS} accurately predicts the shape of the size distribution and the growth distribution at any level of aggregation of economic systems. In this paper we derive the implications of the model on the size-variance relationship. In principle, the predictions of the model can be studied analytically, however due to
the complexity of the resulting integrals and series which cannot be expressed in elementary functions we will rely in our study on
computer simulations. The main conclusion is that the relationship between the size and the variance of growth rates is not a true power law with a single well-defined exponent $\beta$ but undergoes a slow crossover from $\beta=0$ for $S\to 0$ to $\beta=1/2$ for $S\to\infty$.

\section{The Model}
\label{sec:Model}

We model business firms as classes consisting of a random number of units of variable size. The number of units $K$ is defined as in the Simon model \cite{Simon55}. The size of the units $\xi$ evolves according to a geometric brownian motion or Gibrat process \cite{Gibrat31}. 

%For the most natural assumption of the Pure Gibrat process
%f%P_\xi(\xi_i)={\frac{1}{\sqrt{2\pi V_\xi}}}\,\,{\frac{1}{\xi_i}}
%\,\,\exp\left(-(\ln\xi_i-m_\xi)^2/2V_\xi\right),
%\label{e:Pxi}
%\end{equation}or the sizes of the products these distributions are lognormal:
%\begin{equation}

%%%%%% online materials

As in the Simon model, business firms as classes consisting of a random
number of units~\cite{Simon77,Sutton98,DeFabritiis03,Amaral98}.
Firms grow by capturing new business opportunities and the probability that a new opportunity is assigned
to a given firm is proportional to the number of opportunities it
has already got. At each time $t$ a new opportunity is assigned.

With probability $b$, the new  opportunity is taken up by a new firm, so that the average number of firms at time $t$ is $N(t)=N(0)+bt$.

With probability $1-b$, the new opportunity is captured by an active firm $\alpha$ with probability $P_{\alpha}=(1-b)K_{\alpha}(t)/t$, where $K_{\alpha}(t)$
is the number of units of firm $\alpha$ at time $t$.

In the absence of the entry of new firms ($b=0$) the probability distribution of the number of the units in the firms at
large $t$, i.e. the distribution P(K) is exponential:
\begin{equation}
P(K)\approx \frac{1}{K(t)}\exp(-K/K(t)),
\label{P_K_old}
\end{equation}
where $K(t)=[n(0)+t]/N(0)$ is the average number of units in the classes, which linearly grows with time.

If $b>0$, $P(K)$ becomes a Yule distribution which behaves as a power
law for small $K$:
\begin{equation}
P_{new}(K)\sim K^{-\varphi},
\end{equation}
where $\varphi=2+b/(1-b)\ge 2$, followed by the exponential decay of Eq.
(\ref{P_K_old}) for large $K$ with $K(t)=[n(0)+t]^{1-b}n(0)^b/N(0)$ \cite{Simon77,Kazuko}. This model can 
be generalized to the case when the units are born at
any unit of time $t'$ with probability $\theta$, die with
probability $\lambda$, and in addition a new class consisting of one
unit can be created with probability $b'$ by letting
$t=t'(\theta-\lambda+b')$ and probability
$b=b'/(\theta-\lambda+b')$.

In the Simon model opportunities are assumed to be of unit
size so that $S_{\alpha}(t) = K_{\alpha}(t)$.
On the contrary we assume that each opportunity has randomly determined but finite size.
In order to capture new opportunities firms launch new products, open up new
establishments, divisions or units. Each opportunity is taken up by exactly one firm and
the size of the firm is measured by the sum of the sizes of the opportunities it has taken up.
Fig.~\ref{schematic} provides a schematic representation of the model.

In the following we consider products as the relevant
constituent parts of the companies and measure their size in terms of sales. The model can be applied to alternative decompositions of economic systems in relevant subunits (i.e. plants) and measures
of their sizes (i.e. number of employees).

At time $t$, the size of each product $\xi_i(t)>0$ is decreased or
 increased by a random factor $\eta_i(t)>0$ so that
\begin{equation}
\xi_i(t)=\xi_i(t-1)\,\eta_i(t),
\end{equation}
where $\eta_i(t)$, the growth rate of product $i$, is independent random
variable taken from a distribution $P_\eta(\eta_i)$, which has finite mean and standard deviation.
We also assume that $\ln \eta_i$ has finite mean
$\mu_{\eta}\equiv\langle\ln\eta_i\rangle$ and variance
$V_{\eta}\equiv\langle(\ln\eta_i)^2\rangle-\mu_{\eta}^2$.

Thus at time $t$ a firm $\alpha$ has $K_{\alpha}(t)$ products of
size $\xi_i(t)$, ${i=1,2,...K_{\alpha}(t)}$ so that its total size is
defined as the sum of the sales of its products $S_{\alpha}(t)\equiv \sum_{i=1}^{K_{\alpha}}\xi_i(t)$ and its
growth rate is measured as $g=\log(S_{\alpha}(t)/S_{\alpha}(t-1))$.

The probability distribution of firm growth rates $P(g)$ is given by
\begin{equation}
P(g) \equiv \sum_{K=1}^{\infty}P(K)P(g|K),
\label{P_g_sum}
\end{equation}

where $P(g|K)$ is the distribution of the growth rates for a firm
consisting of $K$ products. Using central limit theorem, one can show
that for large $K$ and small $g$, $P(g|K)$
converge to a Gaussian distribution
\begin{equation}
P(g|K) \approx{\frac{\sqrt K}{\sqrt{2\pi V}}}\,\exp\left(-\frac{(g-m)^2K}{2V}\right),
\label{P_g_large_K}
\end{equation}

where $V$ and $m$ are functions of the distributions $P_\xi$ and
$P_\eta$. For the most natural assumption of the Pure Gibrat process
for the sizes of the products these distributions are lognormal:
\begin{equation}
P_\xi(\xi_i)={\frac{1}{\sqrt{2\pi V_\xi}}}\,\,{\frac{1}{\xi_i}}
\,\,\exp\left(-(\ln\xi_i-m_\xi)^2/2V_\xi\right),
\label{e:Pxi}
\end{equation}
\begin{equation}
P_\eta(\eta_i)={\frac{1}{\sqrt{2\pi V_\eta}}}\,\,{\frac{1}{\eta_i}}
\,\,\exp\left(-(\ln\eta_i-m_\eta)^2/2V_\eta\right).
\end{equation}
In this case, \begin{equation}
m = m_\eta+ V_\eta/2
\label{e.m}
\end{equation}
and
\begin{equation}
V\equiv K\sigma^2 = \exp(V_\xi)(\exp(V_\eta)-1),
\label{e.V}
\end{equation}

but for large $V_\xi$ the convergence to a Gaussian is an extremely slow
process. Assuming that the convergence is achieved, one can analytically
show \cite{Fu_PNAS} that $P(g)$ has similar behavior to the Laplace distribution for
small $g$ i.e. $P(g)\approx \exp(-\sqrt{2}|g|/\sqrt{V})/\sqrt{2V}$,
while for large $g$ $P(g)$ has power law wings $P(g)\sim g^{-3}$ which
are eventually truncated for $g\to \infty$ by the distribution $P_{\eta}$
of the growth rate of a single product.

To derive the size variance relationship we must compute 
the conditional probability density of the growth rate $P(g|S,K)$, of an economic system with $K$ units and size $S$. For $K \to
 \infty$ the conditional probability density function $P(g|S,K)$ develops 
a tent shape functional form, because in the center it
converges to a Gaussian distribution with the width decreasing inverse
proportionally to $\sqrt{K}$, while the tails are governed by the
behavior of the growth distribution of a single unit which remains to be wide independently of $K$.
%(Fig. \ref{f:PgSK}).

We can also compute the conditional probability $P(S|K)$, which is
the convolution of $K$ unit size distributions $P_\xi$. In case of lognormal
$P_\xi$ with a large logarithmic variance $V_\xi$ and mean $m_\xi$, the convergence
of $P(S|K)$ to a Gaussian is very slow (see Chapter II).
Since $P(S,K)=P(S|K)P(K)$, we can find
\begin{equation}
P(g|S)=\sum P(g|S,K)P(S|K)P(K),
\label{PgSK}
\end{equation}
where all the distributions $P(g|S,K)$, $P(S|K)$, $P(K)$ can be
found from the parameters of the model. $P(S|K)$ has a sharp maximum
near $S=S_K \equiv K_{\mu_{\xi}}$, where $\mu_\xi=\exp (m_\xi+V_\xi/2)$ is the mean
of the lognormal distribution of the unit sizes. Conversely,
$P(S|K)$ as function of $K$ has a sharp maximum near
$K_S=S/\mu_\xi$. For the values of $S$ such that $P(K_S)>>0$,
$P(g|S)\approx P(g|K_S)$, because $P(S|K)$ serves as a
$\delta(K-K_S)$ so that only terms with $K\approx K_S$ make a
dominant contribution to the sum of Eq. (\ref{PgSK}). Accordingly,
one can approximate $P(g|S)$ by $P(g|K_S)$ and $\sigma(S)$ by
$\sigma(K_S)$. However, all firms with $S<S_1=\mu_\xi$ consist
essentially of only one unit and thus
\begin{equation}
\sigma(S)=\sqrt{V_\eta}
\label{e:smallS}
\end{equation}
for $S<\mu_\xi$. For large $S$ if $P(K_S)>0$
\begin{equation}
\sigma(S)=\sigma(K_S)=\sqrt{V/K_S}={\exp(3V_\xi/4+m_\xi/2)\sqrt{\exp(V_\eta)-1}\over \sqrt{S}}
\label{e:largeS}
\end{equation}
where $m_\eta$ and $V_\eta$ are the logarithmic mean and variance of the 
unit growth distributions $P_\eta$ and $V=\exp(V_\xi)[\exp(V_\eta)-1]$.
Thus one expects to have a crossover from $\beta=0$ for $S<\mu_\xi$
to $\beta=1/2$ for $S>>S^\ast$, where
\begin{equation}
S^\ast=\exp(3V_\xi/2+m_\xi)(\exp(V_\eta)-1)/V_\eta
\label{Sstar}
\end{equation}
is the value of $S$ for which Eq.(\ref{e:smallS}) and  Eq.(\ref{e:largeS}) give
the same value of $\sigma(S)$. Note that for small $V_\eta<1$,
$S^\ast \approx \exp(3V_\xi/2+m_\xi)$. The range of crossover extends from
$S_1$ to $S^\ast$, with $S^\ast/S_1=\exp(V_\xi)\to \infty$ for
$V_\xi\to\infty$. Thus in the double logarithmic plot of $\sigma$
vs. $S$ one can find a wide region in which the slope
$\beta$ slowly vary from 0 to 1/2 ($\beta\approx 0.2$) in agreement with many empirical observations.

The crossover to $\beta=1/2$ will be observed only if
$K^\ast=S^\ast/\mu_\xi=\exp(V_\xi)$ is such that $P(K^\ast)$ is
significantly larger than zero. For the distribution $P(K)$ with a
sharp exponential cutoff $K=K_0$, the crossover will be observed
only if $K_0>>\exp(V_\xi)$.

Two scenarios are possible for $S>S_0=K_0\mu_\xi$. In the first, there will be no economic system with $S>>S_0$. 
In the second, if the distribution of the size of units
$P_\xi$ is very broad, large economic systems can exist just because the size of a unit can be larger
than $S_0$. In this case exceptionally large systems might 
consist of one extremely large unit $\xi_{\rm max}$,
whose fluctuations dominate the fluctuations of the entire system.

One can introduce the effective number of units in a system
$K_e=S/\xi_{\rm max}$, where $\xi_{\rm max}$ is the largest unit of
the system. If $K_e<2$, we would expect that $\sigma(S)$ will again
become equal to its value for small $S$ given by Eq.
({\ref{e:smallS}), which means that under certain conditions
$\sigma(S)$ will start to increase for very large economic systems and
eventually becomes the same as for small ones.

Whether such a scenario is possible depends on the complex interplay
of $V_\xi$ and $P(K)$. The crossover to $\beta=1/2$ will be seen
only if $P(K>K^\ast)>P(\xi>S^\ast)$ which means that such large
systems predominantly consist of a large number of units. Taking into account the equation of $P_\xi$,
one can see that $P(\xi>S^\ast)\sim\exp(-9/8 V_\xi)$. 

On the one hand, for an exponential $P(K)$, this implies that

\begin{equation}
\exp(-\exp(V_\xi)/K_0)>\exp(-9/8 V_\xi)
\end{equation}

or

\begin{equation}
V_\xi>8\exp(V_\xi)/(9 K_0).
\label{CondExp}
\end{equation}

This condition is easily violated if $V_\xi>>\ln K_0$. Thus for the distributions $P(K)$
with exponential cut-off we will never see the crossover to $\beta=1/2$ if $V_\xi >>\ln K_0$.

On the other hand, for a power law distribution $P(K)\sim K^{-\phi}$, the
condition of the crossover becomes $\exp(V_\xi)^{1-\phi}>\exp(-9/8 V_\xi)$, or
$(\phi-1)V_\xi<9/8 V_\xi$ which is rigorously satisfied for

\begin{equation}
\phi < 17/8
\label{CondPL}
\end{equation}

but even for larger $\phi$ is not dramatically
violated. Thus for power law distributions, we expect a
crossover to $\beta=1/2$ for large $S$ and significantly large
number $N$ of economic entities in the data set: $N P(K^\ast)>1$. The sharpness
of the crossover mostly depends on $V_\xi$. For power law
distributions we expect a sharper crossover than for exponential
ones because the majority of the economic systems in a power law distribution have a small number of
units $K$, and hence $\beta=0$ almost up to $S^\ast$, the size at which the
crossover is observed. For exponential distributions we expect a
slow crossover which is interrupted if $V_\xi$ is comparable to $\ln
K_0$. For $S>>S_1$ this crossover is well represented by the
behavior of $\sigma(K_S)$.

We confirm these heuristic arguments by means of computer simulations.
Figure~\ref{f:sigma-all1} shows the behavior of $\sigma(S)$ for the exponential distribution $P(K)=\exp(-K/\langle K\rangle)/\langle K\rangle $
and lognormal $P_\xi$ and $P_\eta$.
We show the results for $K_0=1,10,100,1000,10000$ and $V_\xi=1,5,10$.
The graphs $\sigma(K_S)$ and the asymptote given by
Eq.(\ref{e:largeS}) are also given to illustrate our theoretical
considerations. One can see that for $V_\xi=1$, $\sigma(S)$ almost
perfectly follows $\sigma(K_S)$ even for $\langle K \rangle=10$. However for $V_\xi=5$, the
deviations become large and $\sigma(S)$ converges
to $\sigma(K_S)$ only for $\langle K \rangle>100$. For $V_\xi=10$ the convergence is never achieved.

Figures ~\ref{f:exp} and ~\ref{sigma-all} illustrate the importance of the effective number of units
$K_e$. When $K_S$ becomes larger than $K_0$, $\sigma(S)$ starts to follow $\sigma (K_e)$. Accordingly,
for very large economic systems $\sigma(S)$ becomes almost the same as for small
ones. The maximal negative value of the slope
$\beta_{\rm max}$ of the double logarithmic graphs presented
in Fig. \ref{f:exp}(a) correspond to the inflection points of these graphs, and can be identified
as approximate values of $\beta$ for different values of $K_0$. One
can see that $\beta_{\rm max}$ increases as $K_0$ increases
from a small value close to $0$  for $ K_0 =10$ to
a value close to $1/2$ for $K_0 =10^5$ in agreement with
the predictions of the central limit theorem.

To further explore the effect of the $P(K)$ on the size-variance relationship
we select $P(K)$ to be a pure power law $P(K)\sim K^{-2}$ [Fig.~\ref{f:real}(a)]. Moreover, we consider a realistic
$P(K)$ where $K$ is the number of products by firms in the pharmaceutical industry
 [Fig.~\ref{f:real}(b)]. As we have seen in Chapter II, this distribution
can be well approximated by a Yule distribution with $\phi=2$ and an exponential
cut-off for large $K$. Figure~\ref{f:real} shows that, for a scale-free
power-law distribution $P(K)$, in which the majority of firms are comprised of small number of units, but there is a significant fractions firms comprised of an arbitrary large number of units,
the size variance relationship depicts a steep crossover
from $\sigma = \sqrt{V_\eta}$ given by Eq. (\ref{e:smallS}) for small $S$ to $\sigma =\sqrt{V/K_S}$ given
by Eq. (\ref{e:largeS}) for large $S$, for any value of $V_\xi$ (Riccaboni, 2008).

As we see, the size-variance relationship of economic systems $\sigma(S)$ can
be well approximated by the behavior of $\sigma(K_S)$ [Fig~\ref{f:exp}(a)]. It was shown in Buldyrev (2007) that, 
for realistic $V_\xi$, $\sigma^2(K)$ can be approximated in a wide range of $K$ as $\sigma(K)\sim
K^{-\beta}$ with $\beta \approx 0.2$, which eventually crosses over
to $K^{-1/2}$ for large $K$. In other words, one can write
$\sigma(K) \sim K^{-\beta(K)}$ where $\beta(K)$, defined as the
slope of $\sigma(K)$ on a double logarithmic plot, increases from a
small value dependent on $V_\xi$ at small $K$ to $1/2$ for $K\to
\infty$. Accordingly, one can expect the same behavior for
$\sigma(S)$ for $K_S<K_0$. 

As $K_S$ approaches $K_0$, $\sigma(S)$ starts to deviate from $\sigma(K_S)$ in the upward direction. This
results in the decrease of the slope $\beta(S)$ as $S\to \infty$ and
one may not see the crossover to $\beta=1/2$. Instead, in a quite
large range of parameters $\beta$ can have an approximately constant
value between $0$ and $1/2$.

Thus it would be desirable to derive an exact analytical expression for
$\sigma(K)$ in case of lognormal and independent $P_\xi$ and $P_\eta$. 
% Unfortunately
% the radius of convergence of the expansion of a logarithmic growth rate in inverse
% powers of $K$ is equal to zero, and these expansions have only a formal asymptotic meaning for
% $K\to \infty$. However, these expansions are useful since they demonstrate that $\mu$ and $\sigma$ do
% not depend on $m_\eta$ and $m_\xi$ except for the leading term in $\mu$: $m_0=m_\eta+V_\eta/2$.
Using the fact that the $n$-th moment of the lognormal distribution
\begin{equation}
P_x(x)={\frac{1}{\sqrt{2\pi V_x}}}\,\,{\frac{1}{x}}
\,\,\exp\left(-(\ln x_i-m_x)^2/2V_x\right),
\end{equation}
is equal to
\begin{equation}
\mu_{n,x} \equiv \langle x^n \rangle =\exp(n m_x +n^2 V_x/2)
\end{equation}
we can make an expansion of a logarithmic growth rate in inverse powers
of $K$:
\begin{eqnarray}
g & = &\ln {\sum_{i=1}^K\xi_i\eta_i\over \sum_{i=1}^K\xi_i} \notag \\
& = & \ln \mu_{1,\eta}+ \ln\left( 1+{A \over K(1+B/K)}\right) \notag \\
& = & m_\eta+{V_\eta \over 2}+ {A(1-B/K +B^2/K^2 ...)\over K}
-{A^2(1-B/K+B^2/K^2 ...)^2\over 2K^2} + ... \notag \\
& = & m_\eta+{V_\eta \over 2}+{A\over K} - {AB+A^2/2\over K^2} +O(K^{-3})
\notag
\end{eqnarray}
where
\begin{eqnarray}
A={\sum_{i=1}^K \xi_i(\eta_i-\mu_{1,\eta})\over \mu_{1,\eta}\mu_{1,\xi}} \\
B={\sum_{i=1}^K \xi_i-\mu_{1,\xi}\over \mu_{1,\xi}} .
\end{eqnarray}
Using the assumptions that $\xi_i$, and $\eta_i$ are independent:
$\langle \xi_i \eta_i\rangle= \langle\xi_i\rangle\langle\eta_i\rangle$,
$\langle \eta_i \eta_j\rangle= \langle\eta_i\rangle\langle\eta_j\rangle$,
and $\langle \xi_i \xi_j\rangle= \langle\xi_i\rangle\langle\xi_j\rangle$
for $i\neq j$, we find $\langle A\rangle =0$, $\langle AB\rangle=0$,
$\langle A^2\rangle=CK$, where $C=a(b-1)$ with $a=\exp(V_\xi)$ and
$b=\exp(V_\eta)$. Thus
\begin{eqnarray}
\mu & =  \langle g \rangle & =  \sum_{n=0}^\infty \frac{m_n}{K^{n}} \notag \\
\sigma^2 & =  \langle g^2 \rangle -\mu^2 & =  \sum_{n=1}^\infty \frac{V_n}{K^{n}} ,
\label{e:m_n}
\end{eqnarray}
where $m_0=m_\eta +V_\eta/2$, $m_1=-C/2$, $V_1=C$,
$V_2=C[a(5b+1)/2-1-a^2b(b+1)]$.  The higher terms involve terms like
$\langle A^n \rangle /K^n$, which will become sums of various products
$\langle \xi_i^k(\eta_i-\mu_{1,\eta})^k \rangle$, where $2\leq k \leq
n$.  The contribution from $k=n$ has exactly $K$ terms of
$\mu_{n,\xi}\mu_{1,\xi}^{-n}\sum_{j=0}^n\mu_{j,\eta}\mu_{1,\eta}^{-j}(-1)^{n-j}\binom{j}{n}$
with $\mu_{j,x}\mu_{1,x}^{-j}=\exp(V_x j(j-1)/2)$.  Thus there are
contributions to $m_n$ and $V_n$ which grow as $(ab)^{n(n+1)/2}$ with
$ab>1$, which is faster than the $n$-th power of any $\lambda>0$. Thus the
radius of convergence of the expansions (\ref{e:m_n}) is equal to
zero, and these expansions have only a formal asymptotic meaning for
$K\to \infty$. However, these expansions are useful since they
demonstrate that $\mu$ and $\sigma$ do not depend on $m_\eta$ and
$m_\xi$ except for the leading term in $\mu$: $m_0=m_\eta+V_\eta/2$.

Not being able to derive close-form expressions for $\sigma$, 
we perform extensive computer simulations,
where $\xi$ and $\eta$ are independent random variables taken from
lognormal distributions $P_\xi$ and $P_\eta$ with different $V_\xi$ and
$V_\eta$. The numerical results (Fig.~\ref{f:sigma}) suggest that
\begin{equation}
\ln \sigma^2(K)K/C\approx F_\sigma \left [ \ln (K) - f(V_\xi,V_\eta) \right ],
\end{equation}
where $F_\sigma(z)$ is a universal scaling function describing a
crossover from $F_\sigma(z)\to 0$ for $z\to \infty$ to
$F_\sigma(z)/z \to 1$ for $z \to -\infty$ and  $f(V_\xi,V_\eta)
\approx f_\xi(V_\xi)+f\eta(V_\eta)$ are functions of $V_\xi$ and
$V_\eta$ which have linear asymptotes for $V_\xi\to \infty$ and
$V_\eta \to \infty$ [Fig.~\ref{f:sigma}(b)].

Accordingly, we can try to define $\beta(z)=(1-dF_\sigma/dz)/2$
[Fig.~\ref{f:beta} (a)]. The main curve $\beta(z)$ can be
approximated by an inverse linear function of $z$, when $z\to
-\infty$ and by a stretched exponential as it approaches the
asymptotic value 1/2 for $z \to +\infty$. The particular analytical
shapes for these asymptotes are not known and derived solely from
least square fitting of the numerical data. The scaling
for $\beta(z)$ is only approximate with significant deviations from
a universal curve for small $K$. The minimal value for $\beta$ practically does not depend on $V_\eta$ 
and is approximately
inverse proportional to a linear function of $V_\xi$:
\begin{equation}
\beta_{min} = {1\over p V_\xi +q}\,
\end{equation}
where $p\approx 0.54$ and $q \approx 2.66$ are universal values
[Fig.~\ref{f:beta}(b)].
This finding is
significant for our study, since it indicates that near its minimum,
$\beta(K)$ has a region of approximate constancy with the value
$\beta_{min}$ between 0.14 and 0.2 for $V_\xi$ between 4 and 8. These values
of $V_\xi$ are quite realistic and correspond to the distribution of
unit sizes spanning over from roughly
two to three orders of magnitude (68\% of all units),
which is the case in the majority in economic and ecological systems.
Thus our study provides a reasonable explanation for the abundance of value of
$\beta \approx 0.2$.

The above analysis shows that $\sigma(S)$ is not a true power-law function, but undergoes a crossover
from $\beta=\beta_{min}(V_\xi)$ for small economic systems to $\beta=1/2$ for
large ones. However this crossover is expected only for very broad distributions $P(K)$.
If it is very unlikely to find an economic complex with $K>K_0$, $\sigma(S)$ will
start to grow for $S>K_0\mu_\xi$. Empirical data do not show such an increase (Fig.~\ref{sigma2}),
because in reality there are few giant entities which rely on
few extremely large units. These entities are extremely volatile
and hence unstable. Therefore for real data we do see neither a crossover
to $\beta = 1/2$ nor an increase of $\sigma$ for large economic systems.

\section{The Empirical Evidence}
\label{sec:empirics}

%In this section we test our model at different levels of aggregation
%of economic systems, from the micro level of products to the macro
%level of industrial sectors and national economies. Besides
%pharmaceutical products and firms, we investigate the growth rates
%of all U.S. publicly-traded firms from 1973 to 2004 in all
%industries, based on Security Exchange Commission filings
%(Compustat). Finally, at the macro level, we study the growth rates
%of the gross domestic product (GDP) of $195$ countries from 1960 to
%2004 (World Bank).

Since the size variance relationship depends on the partition of firms into their constituent components,
to properly test our model one must decompose an economic system into parts.
In this section we analyze the pharmaceutical industry database which covers the whole size distribution
for products and firms and monitors flows of entry and exit at
every level of aggregation. Products are classified by companies, markets
and international brand names, with different
distributions $P(K)$ with $\langle K\rangle = K_0$ ranging from 5.8 for international
products to almost 1,600 for markets [Tab.~\ref{SIM}].
If firms have on average $K_0$ products and $V_\xi << \ln K_0 $, the
scaling variable $z=K_0$ is positive and we expect
$\beta \to 1/2$.% [Fig.~\ref{SIM}].
On the contrary, if $V_\xi >>\ln K_0$, $z<0$ and we expect $\beta \to 0$. These
considerations work only for a broad distribution of $P(K)$ with
mild skewness such as an exponential distribution. At the opposite
extreme, if all companies have the same number of products, the
distribution of $S$ is narrowly concentrated near the most probable
value $S_0=\mu_\xi K$ and there is no reason to define $\beta(S)$.
Only very rarely $S>>S_0$, due to a low probability of
observing an extremely large product which dominates the fluctuation
of a firm. Such a firm is more volatile than other firms of equal size.
% as one can see from Fig.~\ref{f:PgSK}.
 This would imply
negative $\beta$. If $P(K)$ is power law distributed, there is a
wide range of values of $K$, so that there are always firms for
which $\ln K >> V_\xi$ and we can expect a slow crossover from
$\beta=0$ for small firms to $\beta=1/2$ for large firms, so that for
a wide range of empirically plausible $V_\xi$, $\beta$ is far form
$1/2$ and statistically different from 0.
The estimated value of the size-variance scaling coefficient $\beta$ goes form $0.123$ for products
to $0.243$ for therapeutic markets with companies in the middle ($0.188$) [Tab.~\ref{SIM} and Fig. ~\ref{sigma-all}].

% supplemental START %%%%%%%%%%%%%%%%%%%%%%%%%%

Our model relies upon general assumptions of independence of the growth
of economic entities from each other and from the number of units $K$. However, these assumptions could be
violated and at least three alternative explanations must be analyzed:

\begin{enumerate}
  \item \textit{Size dependence.} The probability that an active firm captures a new market opportunity
is more or less than proportional to its current size. In particular, there could be a positive relationship
  between the number of products of firm $\alpha$ ($K_{\alpha}$) and
  the size ($\xi_i(\alpha)$) and growth ($\eta_i(\alpha)$) of its component parts
   due to monopolistic effects and economies of scale and scope.
  If large and small companies do not get access to the same distribution of market opportunities, 
large firms can be riskier than small firms simply because they tend to capture bigger opportunities.

  \item \textit{Units interdependence.} The growth processes of the consituent parts of a firm are not independent.
  One could expect product growth rates to be positively
  correlated at the level of firm portfolios, due to product similarities and common management,
  and negatively correlated at the level of relevant markets, due to substitution effects and competition.
  Based on these arguments, one would predict large companies to be less risky than
  small companies because their product portfolios tend to be more diversified.

  \item \textit{Time dependence.} The growth of firms constituent units does not follow a pure Gibrat
  process due to serial auto-correlation and lifecycles. Young products and firms are supposed 
  to be more volatile then predicted by the Gibrat's Law due to
  learning effects. If large firms are older and have more mature products, they should be less risky than small firms.
  On the contrary, ageing and obsolescence would imply that incumbent firms are more unstable than newcomers.

%   \item The frequency of growth shocks is higher at the level of new
%   product arrivals than at the level of product growth shocks.  Our
%   model is based on the assumption that the frequency of growth shocks
%   is higher at lower levels of aggregation and we consider the
%   contribution of $\Delta(K)$ as negligible in the short run but we
%   know that the growth of a firm depends on both product entry/exit as
%   well as the growth of existing products.
\end{enumerate}

The first two hypotheses are not falsified by our data (Fig.~\ref{f:xiK}). 

The number of products of a
firm and their average size defined as $\langle \xi(K) \rangle =\langle {1\over K} \sum_{i=1}^K\xi_i\rangle$, where $\langle \rangle$ indicates averaging over all companies with
$K$ products, has an approximate power law dependence $ \langle
\xi(K) \rangle \sim K^\gamma$, where $\gamma=0.38$. 

The mean correlation coefficient of product growth rates at the firm level $\langle
\rho(K)\rangle$ shows an approximate power law dependence $ \langle
\rho(K) \rangle \sim K^\zeta$, where $\zeta=-0.36$.

Since larger firms are composed by bigger products and are more diversified than small firms the two
effects compensate each other. Thus if products are randomly reassigned to companies, the 
size variance relationship will not change.

As for the time dependence hypothesis, despite there are some departures from a Gibrat process at the product level (Fig.~\ref{f:acorr}) due to lifecycles and seasonal effects, they are too weak to account for the size variance
relationship. Moreover asynchronous product lifecycles are washed out upon aggregation.

% To check first and second candidate explanations, we
%  randomly reassign elementary units to firms,
% markets and molecules. In doing that, we keep the
% number of the products in each class and the history of the fluctuation of each product sales unchanged.
% The origin of a particular value
% of $\gamma$ as well as its universality are the questions which
% require further investigation. However, our simulations reveal that
% the removal of the positive correlation between the product size and
% the number of products does not significantly increase the value of $\beta$. Moreover, it should
% be noticed that even if we assume that the size of products $\xi$ is
% independent form the size of the firm, in real world settings this implies that the
% average size of the products $\langle \xi(K) \rangle$ is independent from
% $K$ only for $K>K_0=200$. In the pharmaceutical industry only the top 3.16 percent
% of the companies that have more than 200 products should in principle have
% the same $\langle \xi(K) \rangle$ (Fig.~\ref{f:xiK}).

% supplemental END %%%%%%%%%%%%%%%%%%%%%%%%%%

To discriminate among different plausible explanaitons we
run a set of experiments in which we keep the real $P(K)$ and randomly
reassign products to firms. In the first
simulation we randomly reassign products by keeping the real world
relationship between the size, $\xi$, and growth, $\eta$, of products. In the second simulation we
reassign also $\eta$. Finally in the last simulation we generate
elementary units according to a geometric brownian motion (Gibrat process) 
with empirically estimated values of the mean and variance of $\xi$ and $\eta$. Tab.~\ref{SIM}
summarizes the results of our simulations.

The first simulation allows us to check for the size dependence and unit interdepence hypotheses
by randomly reassigning elementary units to firms and
markets. In doing that, we keep the
number of the products in each class and the history of the fluctuation of each product sales unchanged.
As for the size dependence, our analysis shows that there
is indeed strong correlation between the number of products in the
company and their average size defined as

\begin{equation}
\langle \xi(K) \rangle =\langle {1\over K} \sum_{i=1}^K\xi_i\rangle,
\label{e:xiK}
\end{equation}

where $\langle \rangle$ indicates averaging over all companies with
$K$ products. We observe an approximate power law dependence $ \langle
\xi(K) \rangle \sim K^\gamma$, where $\gamma=0.38$.
%(Fig.~\ref{f:xiK}).
If this would be a true asymptotic power law holding for $K\to \infty$ than the
average size of the company of $K$ products would be proportional to
$\xi(K)K\sim K^{1+\gamma}$. Accordingly, the average number of
products in the company of size $S$ would scale as $K_0(S) \sim S^{1/(1+\gamma)}$ and consequently due to central limit
theorem $\beta =1/(2+2\gamma)$. In our data base, this would mean
that the asymptotic value of $\beta=0.36$. Similar logic was used to explain $\beta$ in \cite{Amaral97,Bottazzi01}. 
Another effect of random redistribution of units will be the
removal of possible correlations among $\eta_i$ in a single firm (unit interdependence).
Removal of positive correlations would decrease $\beta$, while removal of
negative correlations would increase $\beta$. The mean correlation coefficient
of the product growth rates at the firm level $ \langle
\rho(K) \rangle$ also has an approximate power law dependence $ \langle
\rho(K) \rangle \sim K^\zeta$, where $\zeta=-0.36$. Since larger firms have bigger products and 
are more diversified than small firms the size dependence and unit interdepencence cancel out
and $\beta$ practically does not change if products are randomly reassigned to firms.

%The origin of a particular value
%of $\gamma$ as well as its universality are the questions which
%require further investigation. However, our simulations reveal that
%the removal of the positive correlation between the product size and
%the number of products does not significantly increase the value of $\beta$. Moreover, it should
%be noticed that even if we assume that the size of products $\xi$ is
%independent form the size of the firm, in real world settings this implies that the
%average size of the products $\langle \xi(K) \rangle$ is independent from
%$K$ only for $K>K_0=200$. In the pharmaceutical industry only the top 3.16 percent
%of the companies that have more than 200 products should in principle have
%the same $\langle \xi(K) \rangle$ (Fig.~\ref{f:xiK}).
%
%
\bigskip

\begin{table}[tbh]
\begin{tabular}{lrrrrrr}
\hline
           &         $N$ &   $K_0$ &  $\beta_1$ &  $\beta^{*}_1$ &    $\beta^{*}_2$  &   $\beta^{*}_3$ \\
\hline
 Markets  &        574 &    1,596.9 &  0.243 &       0.213 &     0.232 &     0.221  \\

%Molecules &      5,032 &   182.0 &   0.207 &    0.205 &    0.149 &    0.142 \\

 Firms &      7,184 &    127.5 &  0.188 &    0.196 &    0.125 &    0.127 \\

 International Products &      189,302 &    5.8 &   0.151 &    0.175 &    0.038 &    0.020 \\

 All Products &           916,036 &    --      & 0.123 &     0.123 &    0 &          0  \\
\hline
\end{tabular}
\caption{The size-variance relationship $\sigma(S) \sim S^{-\beta(S)}$: estimated values of $\beta$ and
simulation results $\beta^{*}$ at different levels of aggregations from products to markets. In simulation 1 ($\beta^{*}_1$)
products are randomly reassigned to firms and markets. In simulation 2 ($\beta^{*}_2$) the growth rates of products are reassigned too. In simulation 3 ($\beta^{*}_3$) we reproduce our model with real $P(K)$ and estimated values of
$m_\xi=7.58$ and $V_\xi=2.10$.}
 \label{SIM}
\end{table}

To control the effect of time dependence, we keep the sizes of
products $\xi_i$ and their number $K_\alpha$ at year $t$ for each
firm $\alpha$ unchanged, so
$S_t=\sum_{i=1}^{K_\alpha}\xi_i$ is the same as in the empirical
data. However, to compute the sales of a firm in the following year
$\widetilde{S}_{t+1}=\sum_{i=1}^{K_\alpha}\xi'_i$, we assume that
$\xi'_i=\xi_i\eta_i$, where $\eta_i$ is an annual growth rate of a
randomly selected product. The surrogate growth rate
$\widetilde{g}=\ln\frac{\widetilde{S}_{t+1}}{S_t}$ obtained in this
way does not display any size-variance relationship at the level of
products ($\beta^{*}_2=0$). However, we still observe a size
variance relationship at higher levels of aggregation. This test
demonstrates that $1/3$ of the size variance relationship depends
on the growth process at the level of elementary units which is not
a pure Gibrat process. However, asynchronous product 
lifecycles are washed out upon aggregation and 
there is a persistent size-variance relationship which is 
not due to product auto-correlation.

Finally we reproduced our model with the
empirically observed $P(K)$ and the estimated moments of the lognormal
distribution of products ($m_\xi=7.58$, $V_\xi=4.41$). We generate
$N$ random products according to our model (Gibrat process) with
the empirically observed level of $V_\xi$ and $m_\xi$. 
% In estimating
% $V_\xi$ we considered the size of products in the middle of they
% lifecycles. 
As we can see in Tab.~\ref{SIM}, our model closely reproduce the values of
$\beta$ at any level of aggregation. We conclude that a model of proportional
growth in both the number and the size of economic units
correctly predicts the size-variance relationship and the way it
scales under aggregation.

The variance of the size of the constituent units
of the firm $V_\xi$ and the distribution of units into firms are
both relevant to explain the size variance relationship of firm growth rates.
Simulations results in Fig.~\ref{sigma2} reveal
that if elementary units are of the same size ($V_\xi = 0$) the central limit theorem 
will work properly and $\beta \approx 1/2$. 
As predicted by our model, by increasing
the value of $V_\xi$  we observe at any level of
aggregation the crossover of $\beta$ form $1/2$ to 0. 
The crossover is faster at the level of markets than at
the level of products due to the higher average number of units per class $K_0$.
However, in real world settings the central limit theorem never
applies because firms have a small number of components of variable size
($V_\xi>0$). For empirically plausible values of $V_\xi$ and $K_0$ $\beta \approx 0.2$.
% In all this simulations we keep real world values for $P(K)$ and the
% size and growth distribution of elementary units.
% We have further
% simulations to check the influence of $P(K)$ and of elementary units
% on the size-variance relationship by assuming that $P(K)$ is a delta
% function and $P(K)$ is an exponential distribution with $\langle K
% \rangle=200$, which is close to the empirical value from the
% pharmaceutical database, and a power law distribution $P(K) \sim
% K^{-2}$. In the first case, $\sigma(S)$ is an increasing function of
% $S$, since large classes can be obtained when one of the units is
% exceptionally large. In this case this large unit is by an order of
% magnitude larger than the rest of the units, accordingly, the
% effective number of units in the large classes is smaller than in
% the small classes. In case of exponential distribution, $\sigma(S)$
% has a region of approximate power law with exponent $\beta \approx
% 0.09 $ , however for very large classes, it again starts to increase
% due to the same mechanism as for the fixed $K$. Finally, for the
% power law distribution we have a strong crossover from a very small
% $\beta\approx 0$ for small classes to $\beta=0.5$ for large classes,
% as expected from the central limit theorem. Interestingly, once we
% select the realistic distribution of $K$, which is close to the
% power law with an exponential cutoff predicted by the model, we
% observe a long region of approximately power law behavior with
% $\beta \approx 0.125$. The scaling of the size variance relationship
% crucially depends on $P(K)$.

\section{Discussion}
\label{sec:conclusion}

Firms grow over time as the economic
system expands and new investment opportunities become available.
To capture new business opportunities firms open new plants and launch
new products, but the revenues and return to the investments are uncertain. 
If revenues were independent random variables drawn from a Gaussian distribution with
mean $m_e$ and variance $V_e$ one should expect that the standard deviation of the sales growth rate of
a firm with K products will be $\sigma(S) \sim S^{-\beta(S)}$ with $\beta=1/2$ and $S=m_eK$.
On the contrary, if the size of business opportunities is given by a geometric brownian motion
(Gibrat's process) and revenues are independent random variables drawn from a lognormal
distribution with mean $m_\xi$ and variance $V_\xi$ the central limit theorem
does not work effectively and $\beta(S)$ exhibits a crossover from
$\beta=0$ for $S\to 0$ to $\beta=1/2$ for $S\to\infty$. For realistic distributions of the
number and size of business opportunities, $\beta(S)$ is approximately constant, as it varies in the range
from 0.14 to 0.2 depending on the average number of units in the firm $K_0$ and the variance
of the size of business opportunities $V_\xi$.
This implies that a firm of size $S$ is expected to be
riskier than the sum of $S$ firms of size 1, even in the case of constant returns to scale and
independent business opportunities.

%%%%%%%%%%%%%%%%%%%%%%%%%%%%%%%%%%%END %%%%%%%%%%%%%%%%%%%%%%%%%%%%%%%%%%%%%

\newpage
%    ------MODEL ------------------------------------------

\begin{figure}[htb]
\centering
\includegraphics[scale=.5,angle=-90]{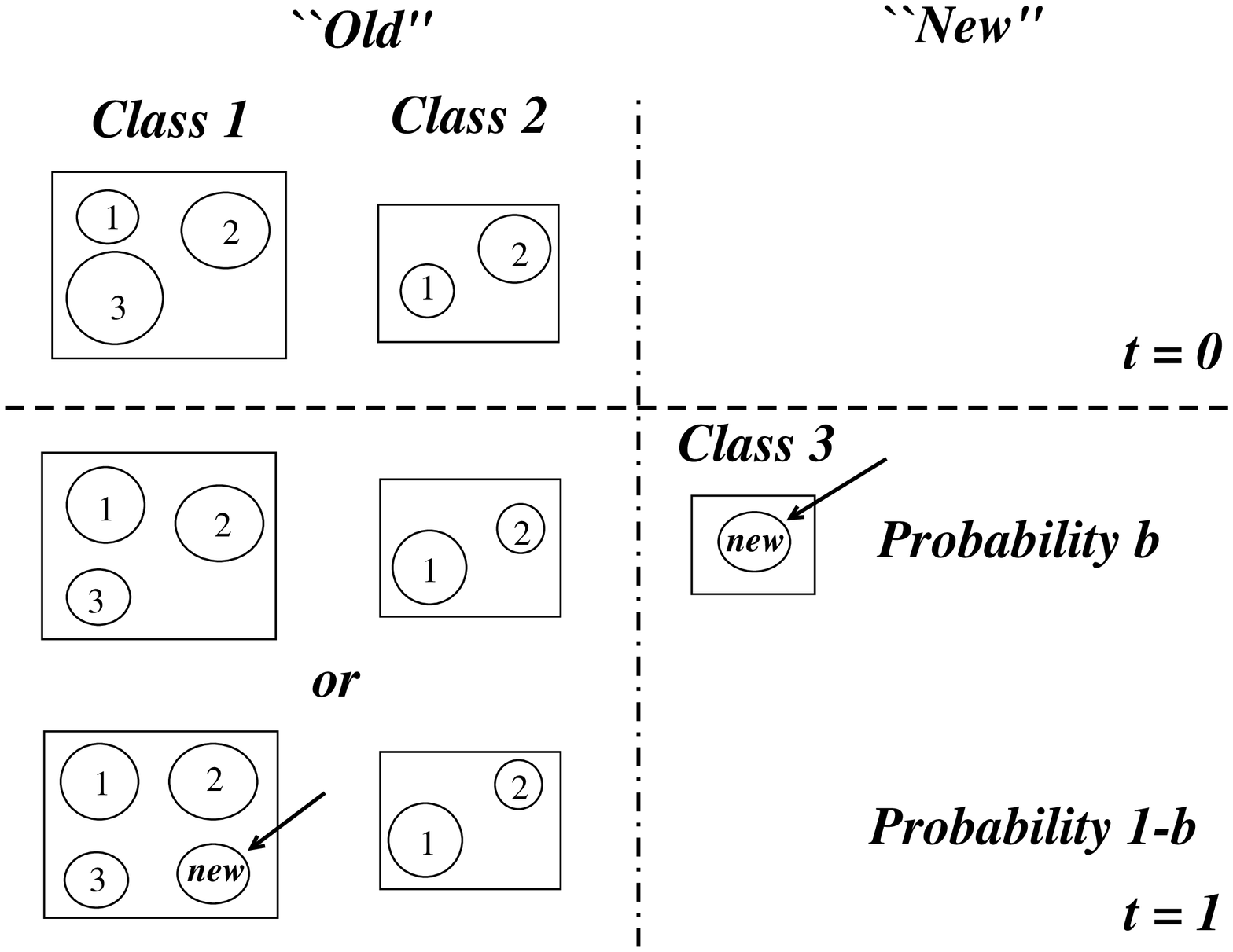}
\caption{Schematic representation of the model of proportional
growth. At time $t=0$, there are $N(0)=2$ classes ($\square$) and
$n(0)=5$ units ($\bigcirc$) (Assumption A1). The area of each circle
is proportional to the size $\xi$ of the unit, and the size of each
class is the sum of the areas of its constituent units (see
Assumption B1). At the next time step, $t=1$, a new unit is created
(Assumption A2). With probability $b$ the new unit is assigned to a
new class (class 3 in this example) (Assumption A3). With
probability $1-b$ the new unit is assigned to an existing class with
probability proportional to the number of units in the class
(Assumption A4). In this example, a new unit is assigned to class
$1$ with probability $3/5$ or to class $2$ with probability $2/5$.
Finally, at each time step, each circle $i$ grows or shrinks by a
random factor $\eta_i$ (Assumption B2).} \label{schematic}
\end{figure}

\begin{figure}
%\begin{center}
\includegraphics[scale=.29,angle=-90]{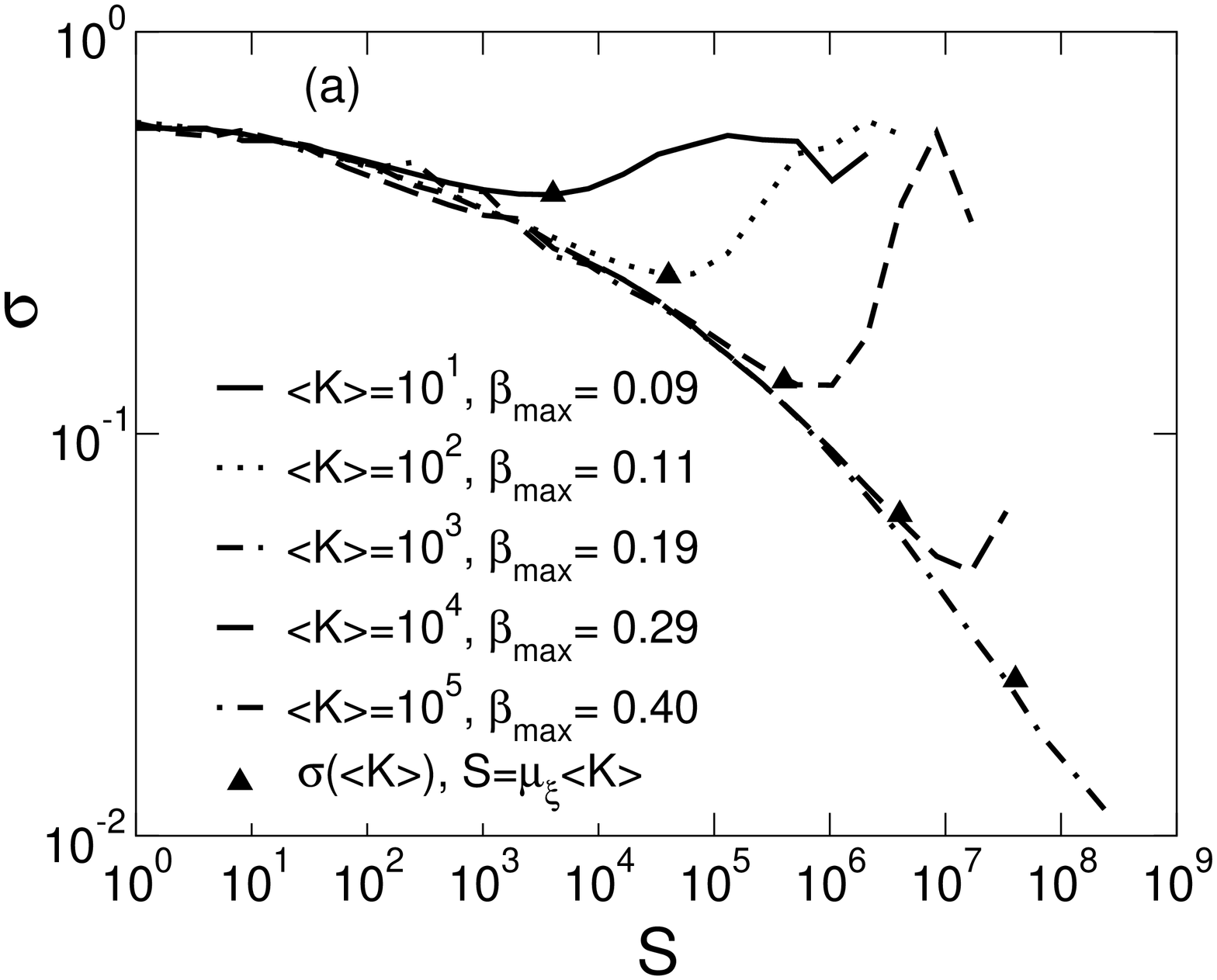}
\includegraphics[scale=.29,angle=-90]{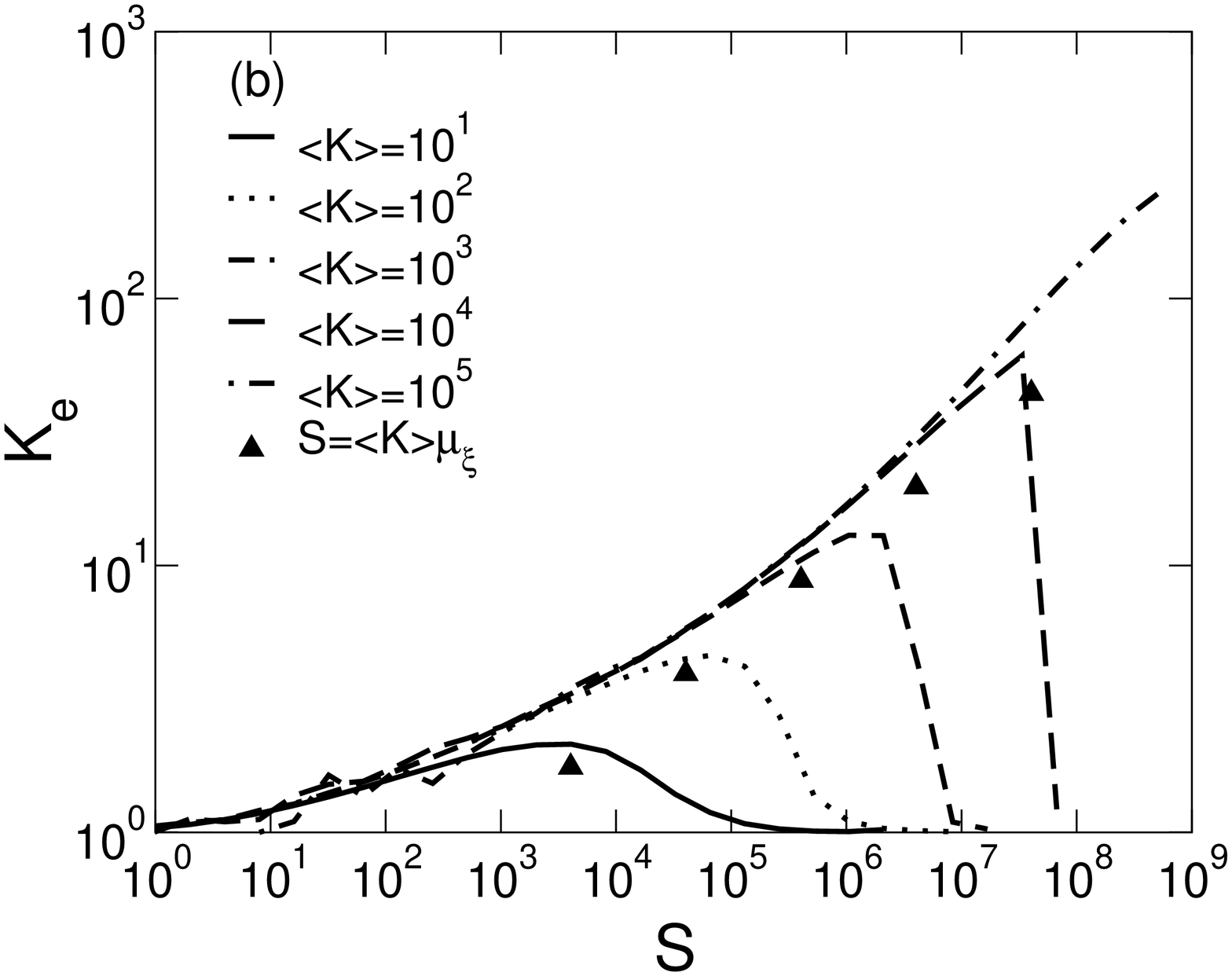}
\caption{(a) Simulation results for $\sigma(S)$ according to Eq. (\ref{PgSK})
for exponential $P(K)=\exp(-K/K_0)/K_0 $ with
$K_0=10,10^2,10^3,10^4,10^5$ and lognormal $P_\xi$ and $P_\eta$
with $V_\xi=5.13, m_\xi=3.44, V_\eta=0.36, \mu_\eta=0.016$ computed for the pharmaceutical database.
One can see that, for small enough $S$ and for different
$K_0$, $\sigma(S)$ follows a universal curve which can be well approximated
with $\sigma(K_S)$, with $K_S=S/\mu_\xi \approx S/405$. For
$K_S>K_0$, $\sigma(S)$ departs from the universal behavior
and starts to increase. This increase can be explained by the decrease
of the effective number of units $K_e(S)$ for the extremely large firms. The maximal negative slope $\beta_{\rm max}$
increases as $K_0$ increases in agreement with
the predictions of the central limit theorem.
(b) One can see, that   $K_e(S)$ reaches its maximum around
 $S\approx K \mu_\xi$.
The positions of maxima in $K_e(S)$ coincide with the positions
of minima in $\sigma(S)$.}.
\label{f:exp}
%\end{center}
\end{figure}

\begin{figure*}
\begin{center}
\includegraphics[scale=.29,angle=-90]{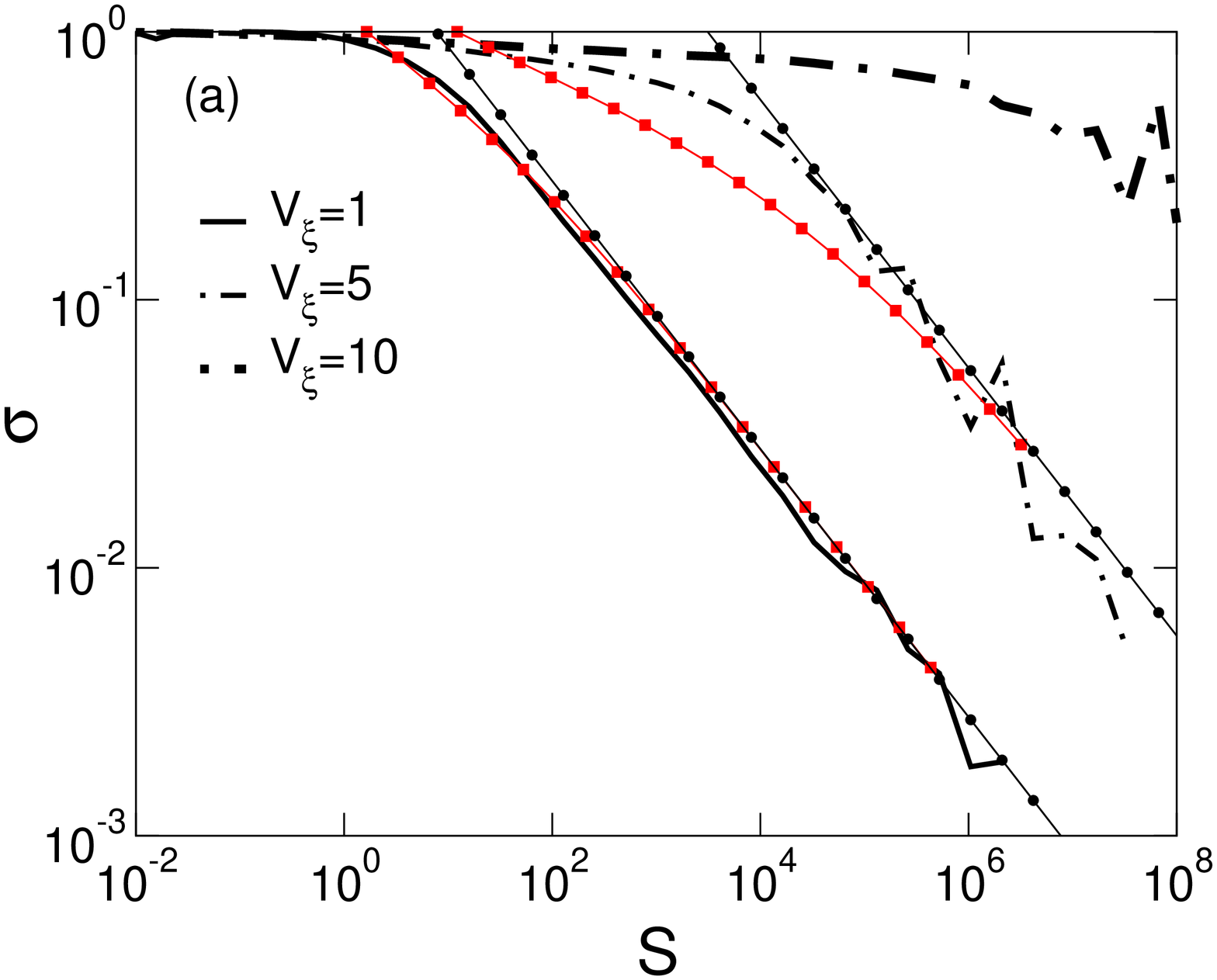}
\includegraphics[scale=.29,angle=-90]{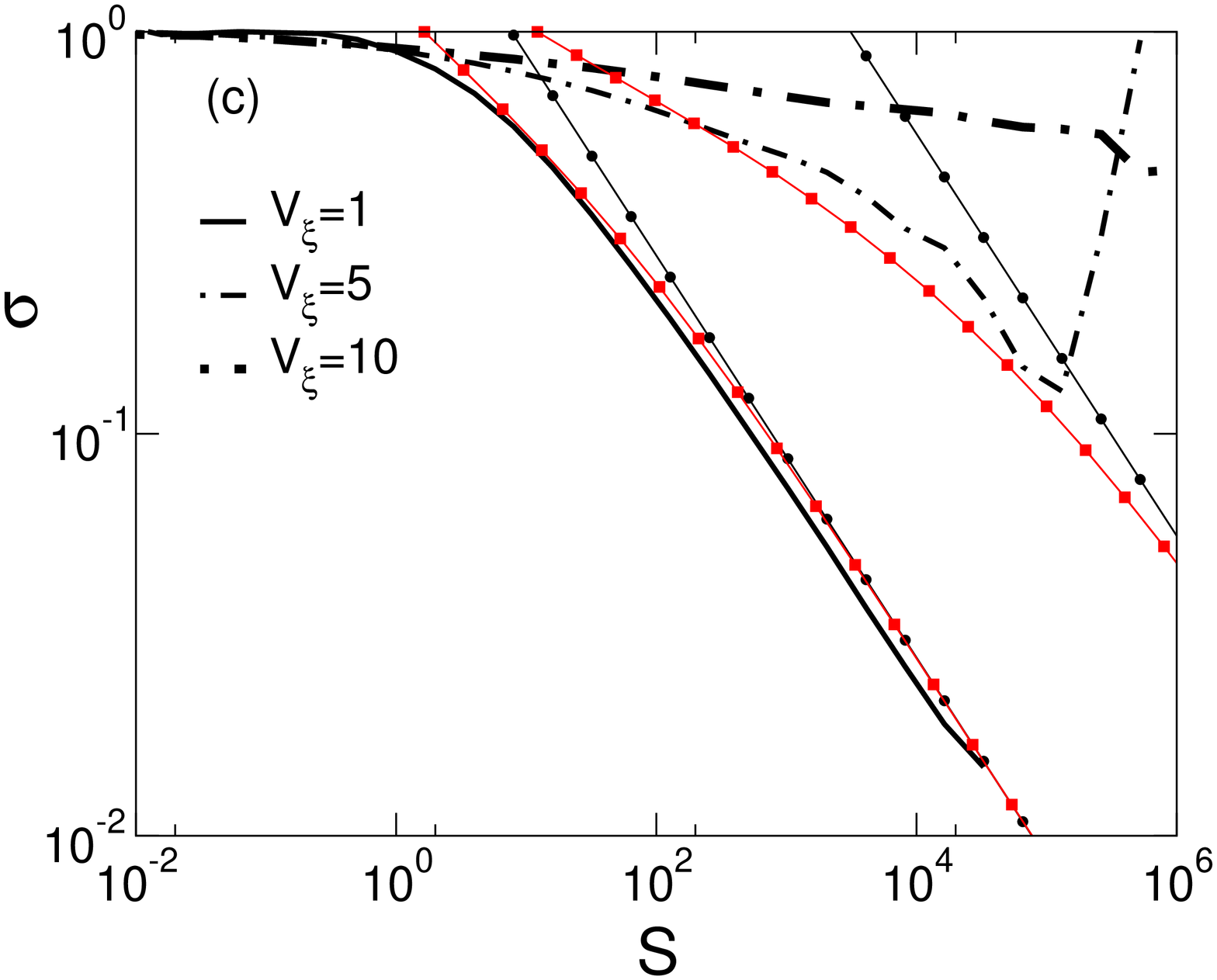}
\caption{%(a) Distributions of the number of products in the
%pharmaceutical database in comparison with a power law distribution $P(K)\sim K^{-2}$.
Size variance relationship $\sigma(S)$ for various $V_\xi$ with $P(K)\sim K^{-2}$ (a) and real $P(K)$ (b).%, and distribution of number of packs (d).
A sharp crossover from $\beta=0$ to $\beta=1/2$ is seen for the power law distribution
even for large values of $V_\xi$. In case of real $P(K)$ one can see a wide crossover
regions in which $\sigma(S)$ can be approximated by a power-law relationship
with $0<\beta<1/2$. Note that the slope of the graphs ($\beta$) decreases with the increase of
$V_\xi$. The graphs of $\beta(K_S)$ and their asymptotes are also shown with squares and circles,
respectively.}
\label{f:real}
\end{center}
\end{figure*}

\begin{figure*}
\begin{center}
\includegraphics[scale=.29,angle=-90]{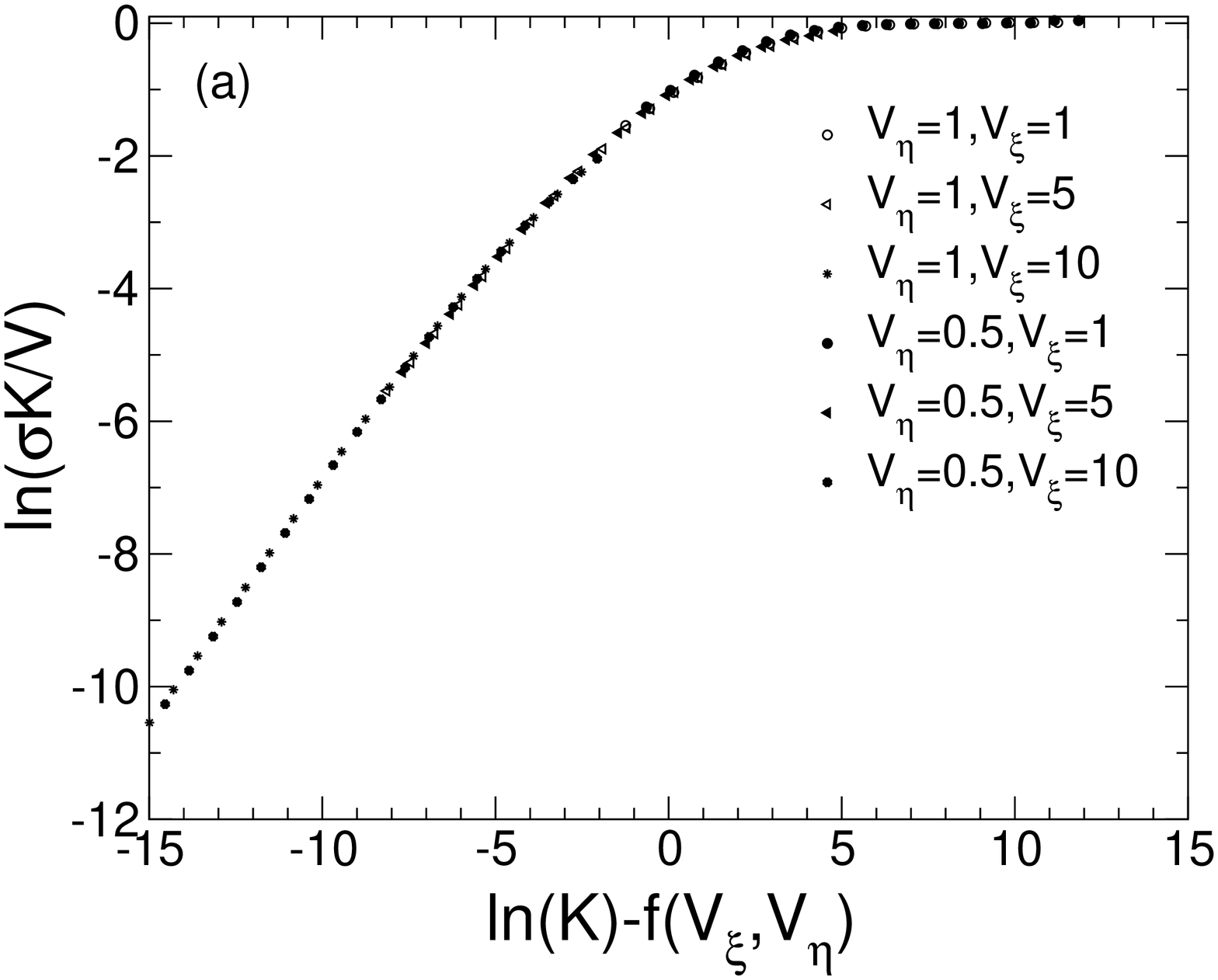}
\includegraphics[scale=.29,angle=-90]{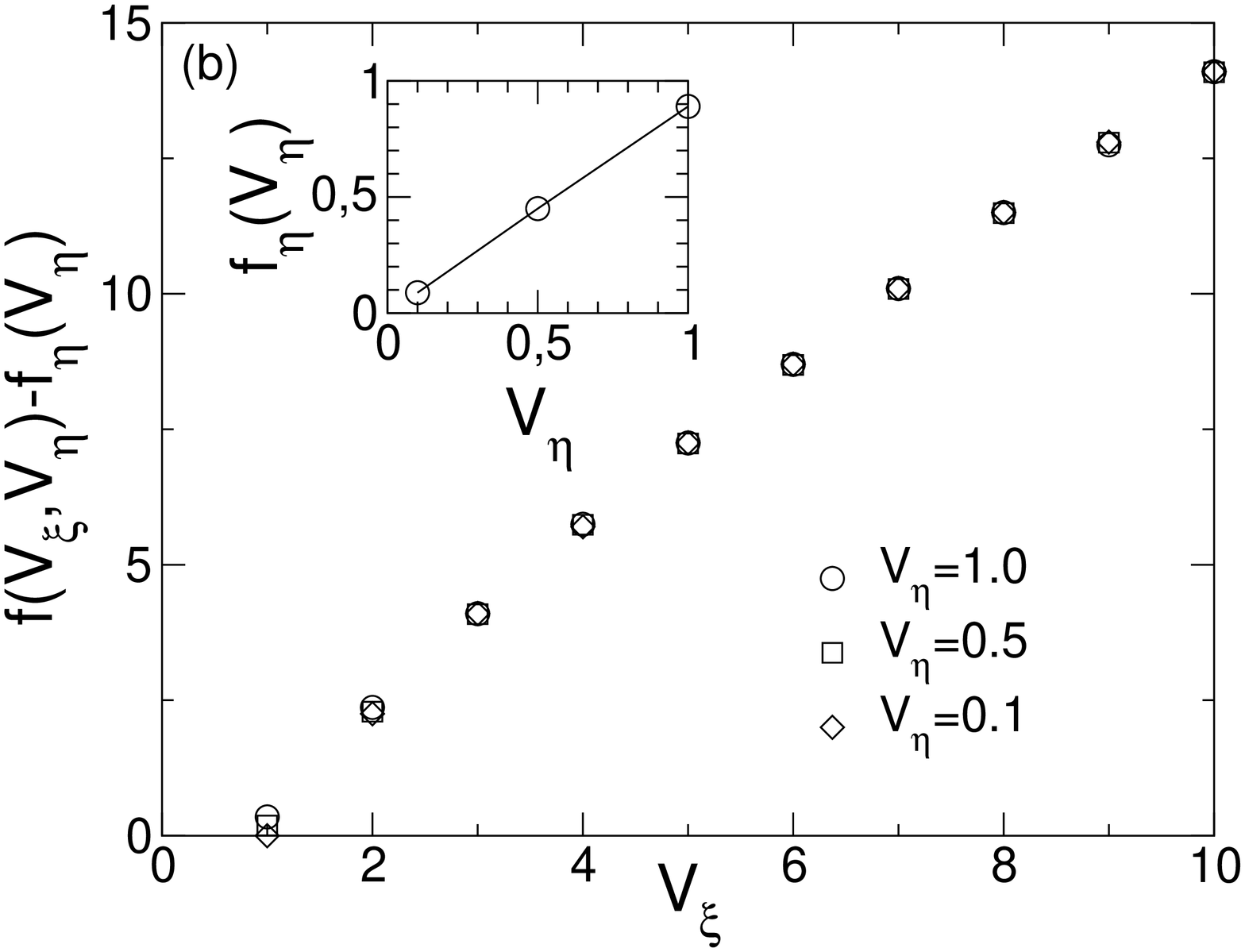}
\caption{(a) Simulation results for $\sigma^2(K)$ in case of lognormal $P_\xi$ and
$P_\eta$ and different $V_\xi$ and $V_\eta$ plotted on a universal
scaling plot as a functions of scaling variable $z=\ln(K)-f(V_\xi,V_\eta)$.
(b) The shift function  $f(V_\xi,V_\eta)$. The graph shows that
$ f(V_\xi,V_\eta)\approx f_\xi(V_\xi)+f_\eta(V_\eta)$ Both
$f_\xi(V_\xi)$ and $f_\eta(V_\eta)$ (inset) are approximately linear
functions.}
\label{f:sigma}
\end{center}
\end{figure*}

\begin{figure*}
\begin{center}
\includegraphics[scale=.29,angle=-90]{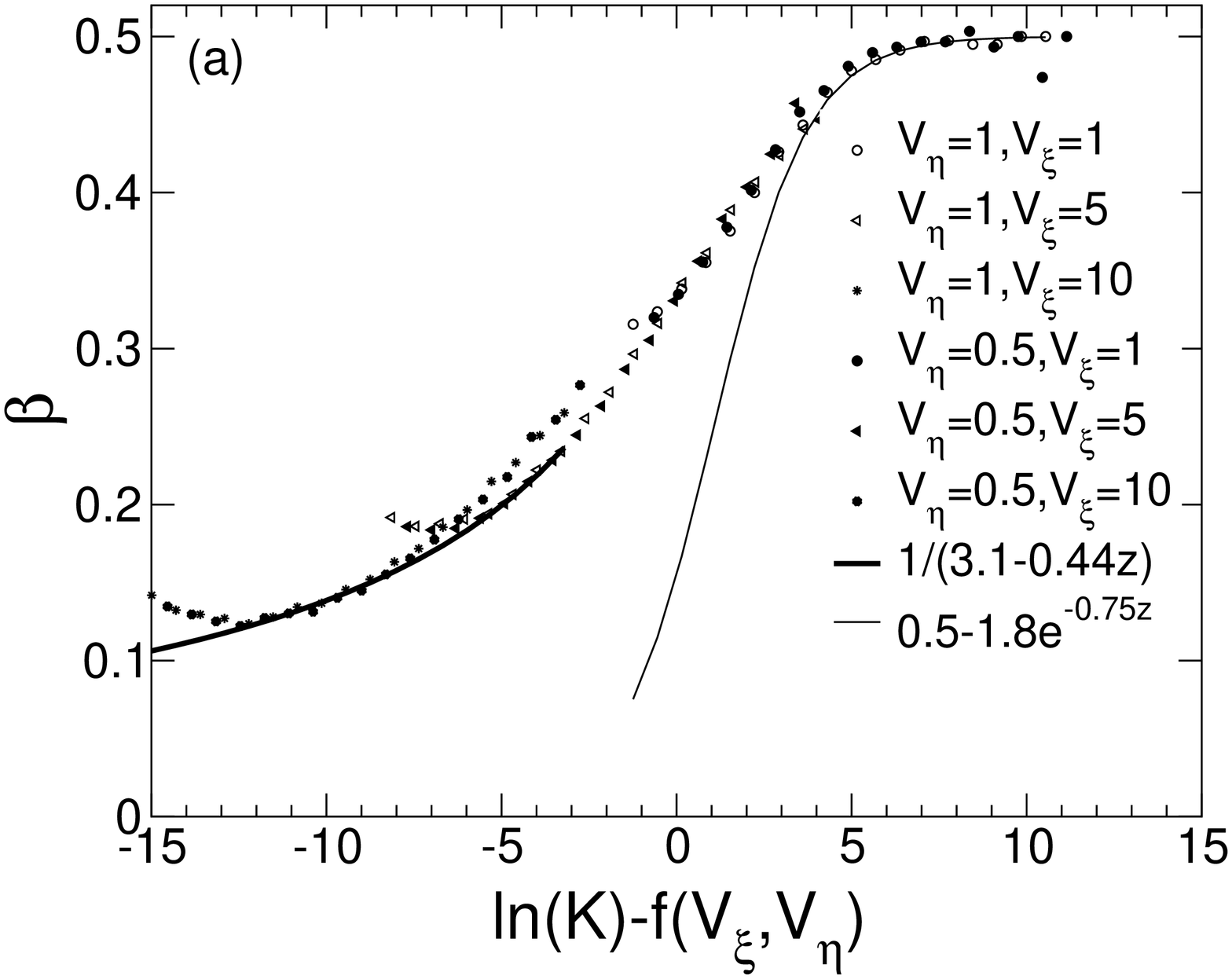}
\includegraphics[scale=.29,angle=-90]{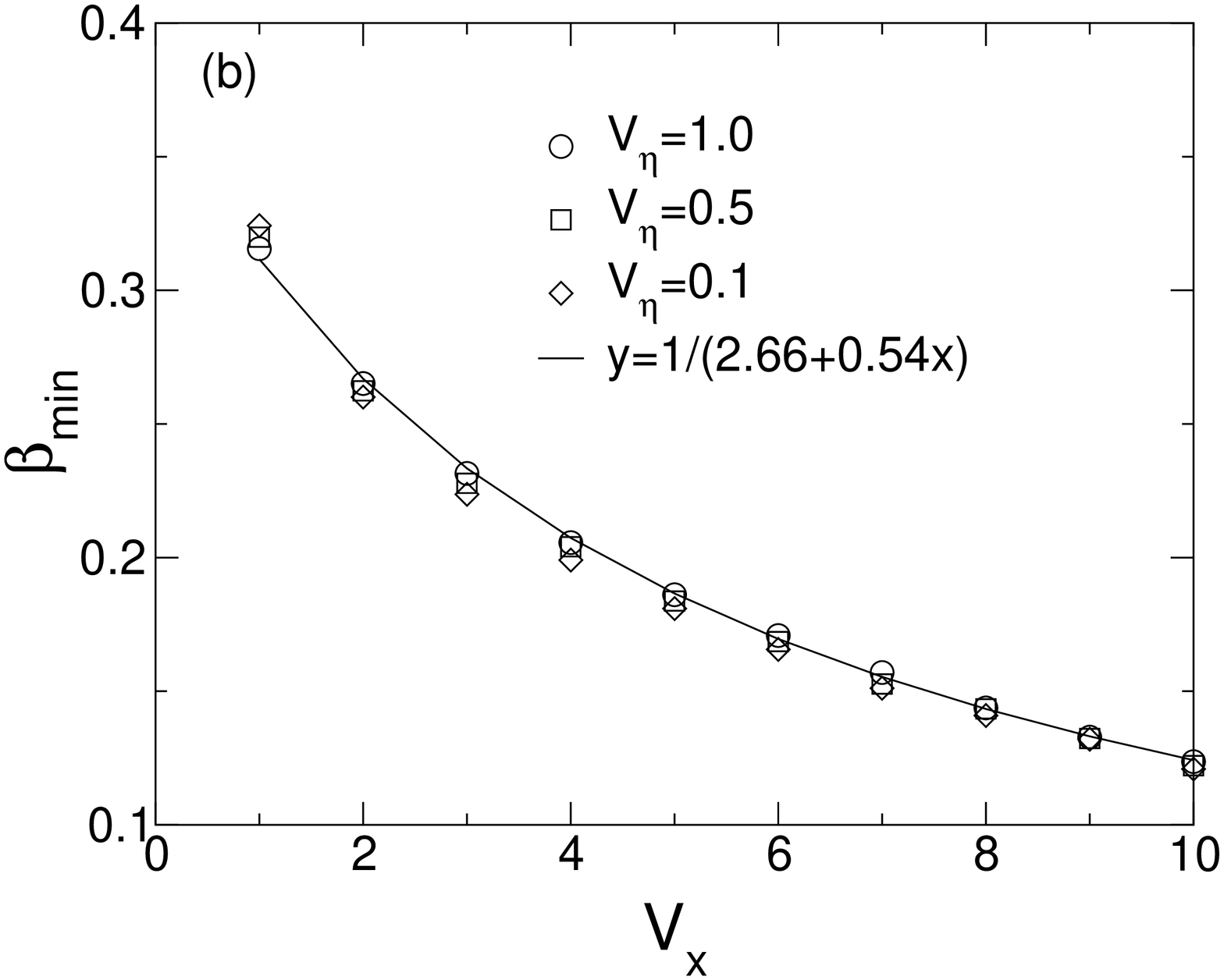}
\caption{(a) The effective exponent $\beta(z)$ obtained by differentiation
of $\sigma^2(z)$ plotted  in Fig.~\ref{f:sigma} (a). Solid lines indicate
least square fits for the left and right asymptotes. The graph shows
significant deviations of $\beta(K,V_\xi,V_\eta)$ from a universal function
$\beta(z)$ for small $K$, where $\beta(K)$ develops minima.
(b) The dependence of the minimal value of $\beta_{min}$ on $V_\xi$.
One can see that this value practically does not depend on $V_\eta$ and
is inverse proportional to the linear function of $V_\xi$.}
\label{f:beta}
\end{center}
\end{figure*}

\begin{figure*}
\begin{center}
\includegraphics[scale=.5,angle=-90]{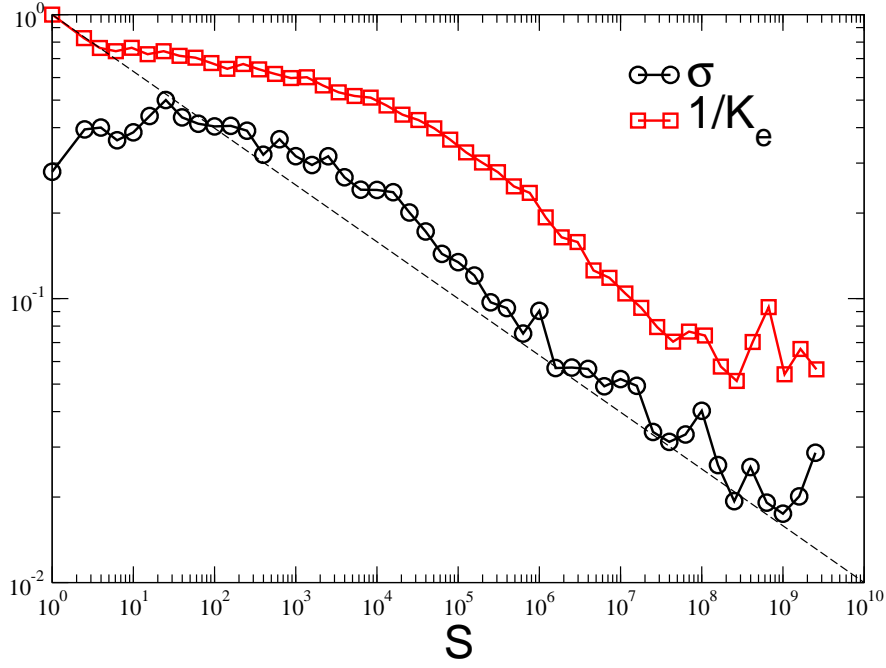}
\caption{The standard error of firm growth rates $(\sigma)$
(circles), and the share of the largest products $(1/K_e)$ (squares)
versus the size of the firm $(S)$. As
predicted by our model for $S<S_1=\mu_\xi\approx 3.44$, $\beta
\approx 0$. For $S>S_1$ $\beta$ increases but never reaches 1/2 due
to the slow grow of the effective number of products $(K_e)$. The
flattering of the upper tail is due to some large companies with
unusually large products.}
\label{sigma-all}
\end{center}
\end{figure*}

\begin{figure*}
\begin{center}
%\centerline{\includegraphics[scale=.3,angle=-90]{chapV/scaling2.ps}}
\includegraphics[scale=.5,angle=-90]{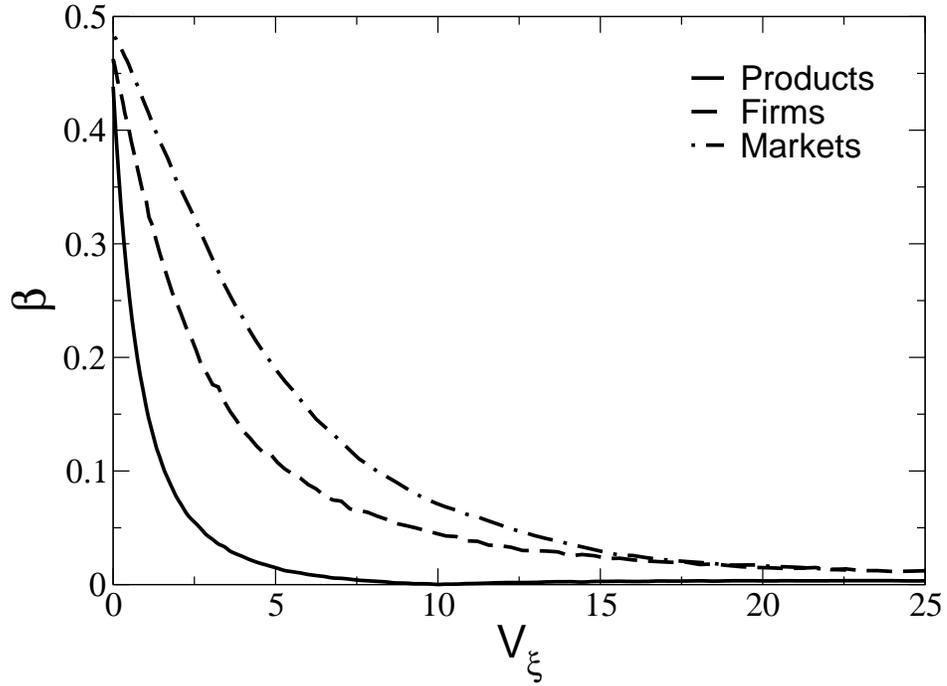}
\caption{The scaling of the size-variance relationship as a function of $V_\xi$.
$\beta$ decays rapidly from $1/2$ to $0$ for $V_\xi \to \infty$. In
the simulation we keep the real $P(K)$ for products, companies and
markets and assign products drawn from a lognormal distribution with
the empirically observed mean $m_\xi$ and variance $0<V_\xi\leq 25$.}
\label{sigma2}
\end{center}
\end{figure*}

\begin{figure*}
\begin{center}
\centerline{\includegraphics[scale=.5,angle=-90]{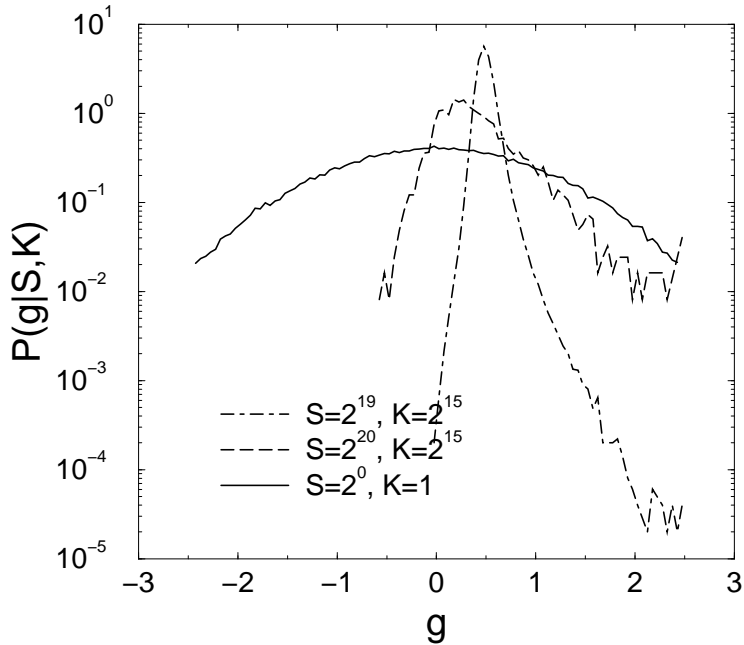}}
\caption{Simulation results for the conditional growth rate
distribution $P(g|S,K)$ for the case of lognormal $P_\xi$ and
$P_\eta$, with $V_\xi=6$, $V_\eta=1$ and $m_\xi=m_\eta=0$. For $K=1$ the distribution is perfectly Gaussian with $V_\eta=1$
and $m_\eta=0$. However for large $K$ the distribution develops a
tent-shape form with the central part close to a Gaussian with mean
$m=1/2$ as predicted by Eq. (\ref{e.m}). The vast majority of
firms (99.7\%) have sizes in the vicinity of $K\mu_\xi$ which for
$K=2^{15}$ and $\mu_\xi=\exp(m_\xi+ V_\xi/2)=20.1$ belongs to the bin $[2^{19},2^{20}]$ and only 0.25\% of firms belong to the next bin $[2^{20},2^{21}]$.
These firms are due to a rare occurrence of extremely large products. The real number of products in these firms is $K_e=2.4$, while the normally sized firms have $K_e=31$.
The fluctuations of these extremely large products dominate the fluctuations of the firm size and
hence $P(g|S,K)$ for such abnormally large firms is broader than for normally
sized firms. Accordingly, $\sigma=0.09$ and $\sigma=0.41$ respectively for the
normally sized and abnormally large firms.}
\label{f:PgSK}
\end{center}
\end{figure*}

\bigskip

\begin{figure*}
\begin{center}
\centerline{\includegraphics[scale=.5,angle=-90]{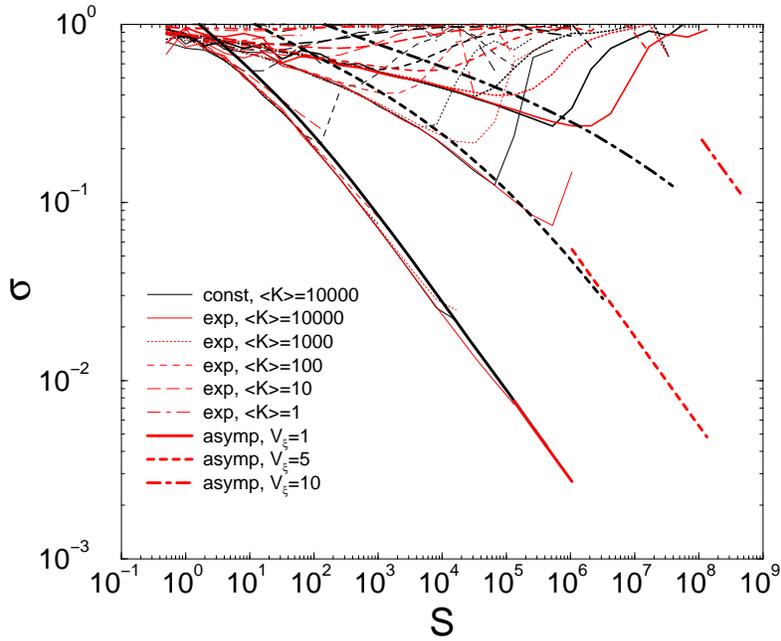}}
\caption{The behavior of $\sigma(S)$ for the exponential distribution $P(K)=\exp(-K/\langle K\rangle)/\langle K\rangle$
 and lognormal $P_\xi$ and $P_\eta$.
We show the results for $K_0=1,10,100,1000,10000$ and $V_\xi=1,5,10$.
The graphs $\sigma(K_S)$ and the asymptote given by $\sigma(S)=\sqrt{V/K_S}={\exp(3V_\xi/4+m_\xi/2)\sqrt{\exp(V_\eta)-1}\over \sqrt{S}}$ are also given to illustrate our theoretical
considerations. One can see that for $V_\xi=1$, $\sigma(S)$ almost
perfectly follows $\sigma(K_S)$ even for $\langle K \rangle=10$. However for $V_\xi=5$, the
deviations become large and $\sigma(S)$ converges
to $\sigma(K_S)$ only for $\langle K \rangle>100$. For $V_\xi=10$ the convergence is never achieved.}
\label{f:sigma-all1}
\end{center}
\end{figure*}

\begin{figure*}[ht]
\begin{center}
\includegraphics[scale=.29,angle=-90]{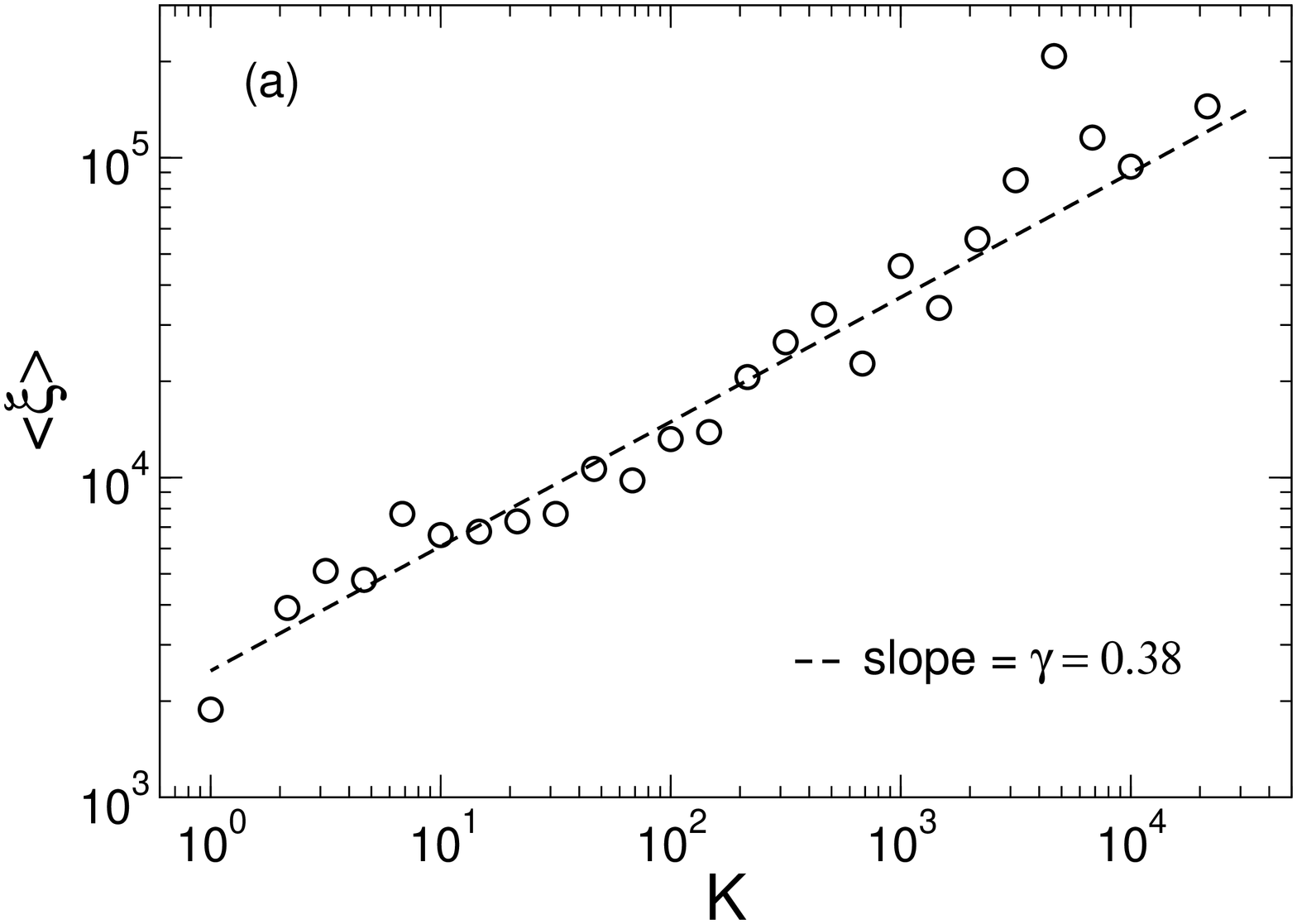}
\includegraphics[scale=.29,angle=-90]{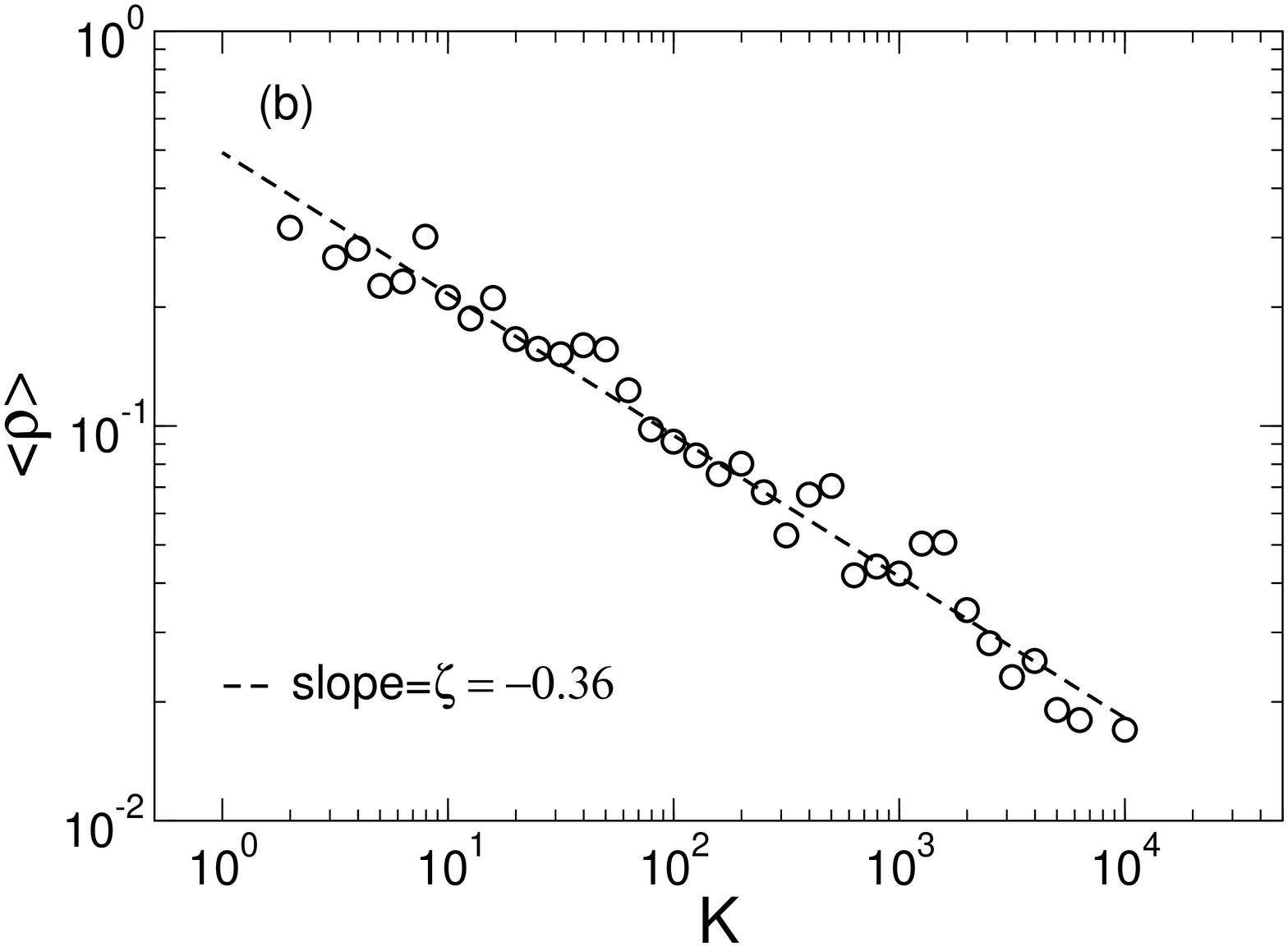}
\caption{(a) The relationship between the average product size and the number
of products of the firm. The log-log plot of $\langle
\xi(K)\rangle$ vs. $K$ shows power law dependence $\langle
\xi(K)\rangle \sim K^{0.38}$. (b) The relationship between the mean correlation coefficient of product growth rates and 
the number of products of a firm. The log-log plot of $\langle
\rho(K)\rangle$ vs. $K$ shows power law dependence $\langle
\rho(K)\rangle \sim K^{0.38}$.}
\label{f:xiK}
\end{center}
\end{figure*}

\begin{figure*}[ht]
\begin{center}
\includegraphics[scale=.29,angle=-90]{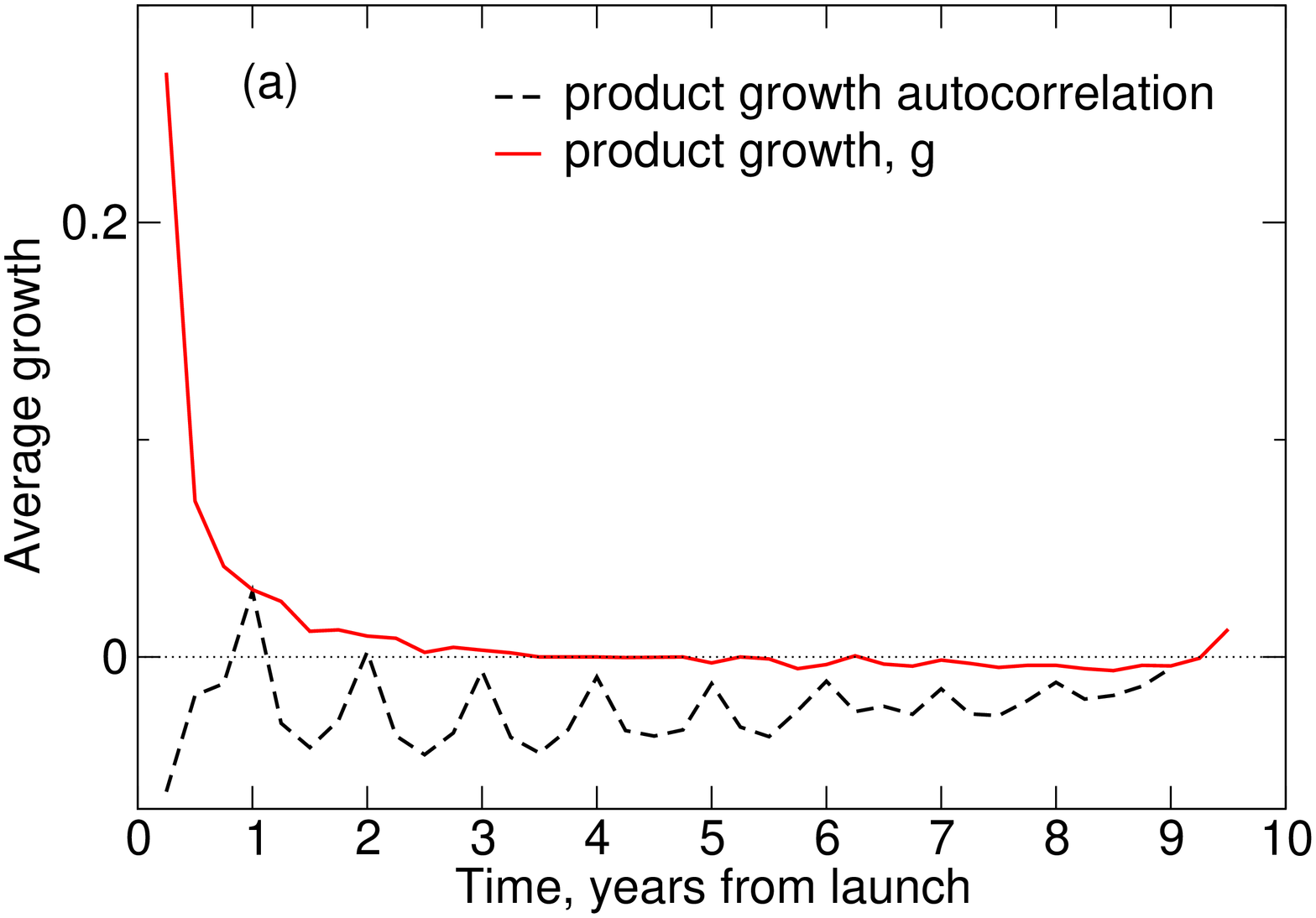}
\includegraphics[scale=.29,angle=-90]{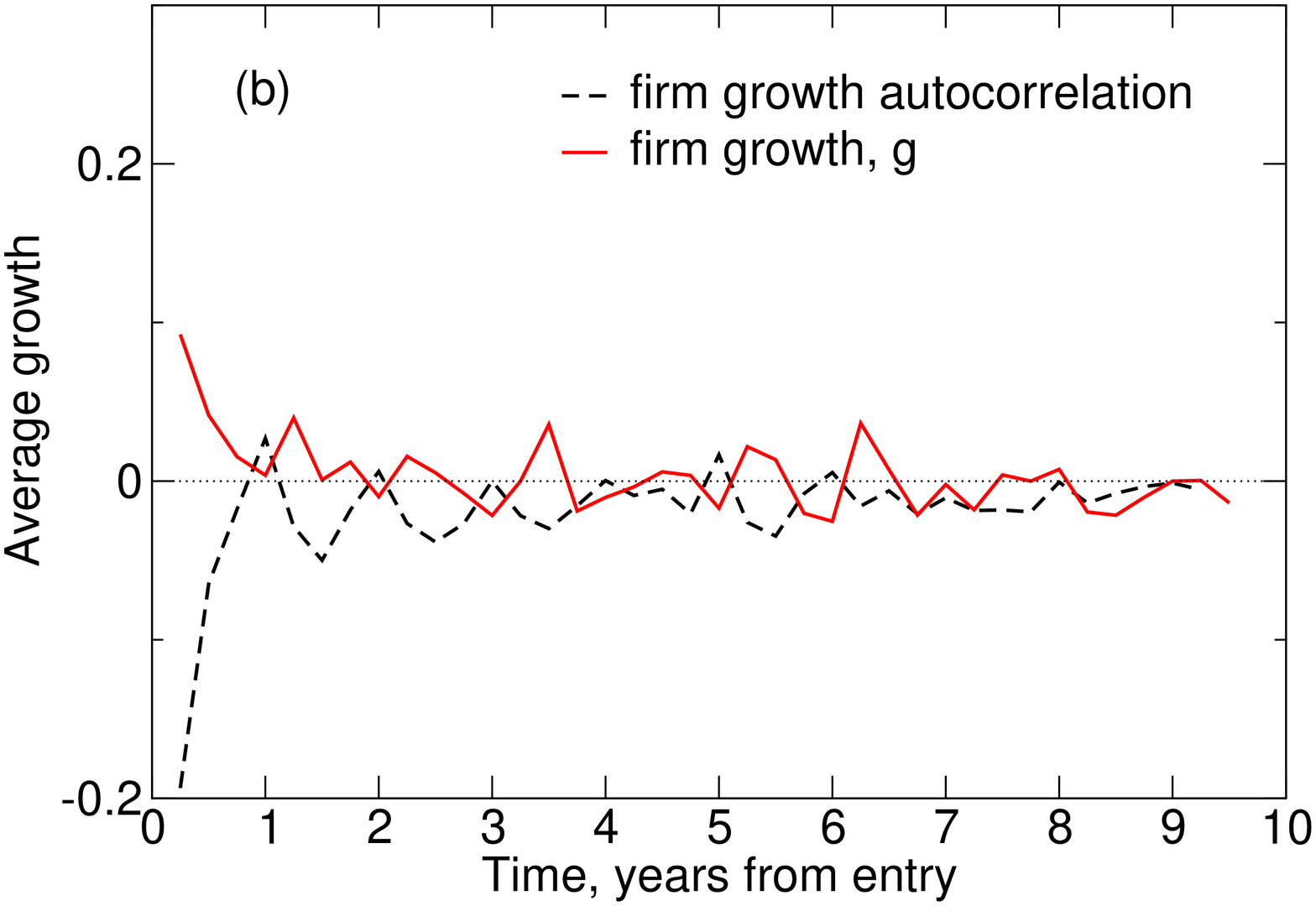}
\caption{(a) The average growth and the auto-correlation coefficient of products since launch. Products growth tend
to be higher in the fist two years from entry. We detect seasonal cycles and a weak (not significant) negative auto-correlation.
(b) The average growth rate and the auto-correlation coefficient of firms from entry. The departures of product growth from a Gibrat process are washed out upon aggregation. The growth rates do not depend on age and do not show a significant auto-correlation.}
\label{f:acorr}
\end{center}
\end{figure*}


\begin{thebibliography}{99}

\bibitem{Gibrat31} Gibrat~R (1931) \textit{Les in\'egalit\'es \'{e}conomiques} (Librairie du Recueil Sirey, Paris).

\bibitem{Sutton97} Sutton~J (1997) {Gibrat's legacy}. {\it J Econ Lit} 35:40-59.

% \bibitem{Kapteyn16} Kapteyn,~J. \& Uven~M.~J. (1916) \textit{Skew Frequency
% Curves in Biology and Statistics} (Hoitsema Brothers, Groningen).
%
% \bibitem{Zipf49} Zipf,~G. (1949) \textit{Human Behavior and the Principle of
% Least Effort} (Addison-Wesley, Cambridge, MA).
%

\bibitem{Kalecki45} Kalecki~M (1945) {On the Gibrat distribution}. \textit{Econometrica} 13:161-170. %On the Gibrat Distribution

\bibitem{Steindl65} Steindl~J (1965) \textit{Random processes and the growth of
firms: a study of the Pareto law} (Griffin, London).

\bibitem{Simon77} Ijiri~Y, Simon~H~A (1977) {\it Skew distributions and the sizes of business firms} (North-Holland Pub. Co., Amsterdam).

\bibitem{Sutton98} Sutton~J, (1998) {\it Technology and market structure: theory and history} (MIT Press, Cambridge, MA.).

%\bibitem{Simon58} Simon,~H.~A. \& Bonini,~C.~P. (1958) {\it Am. Econ. Rev.}
%{\bf 48}, 607-617.

\bibitem{Simon55} Simon~H~A (1955) {On a class of skew distribution functions}. {\it Biometrika} 42:425-440.

\bibitem{Simon75} Ijiri~Y, Simon~H~A (1975) {Some distributions associated with Bose-Einstein statistics}. {\it Proc Natl Acad Sci USA} 72:1654-1657.

\bibitem{Yule25} Yule~U (1925) {A mathematical theory of evolution, based on the conclusions of Dr. J. C. Willis}. \textit{ Philos Trans R Soc London B} 213:21-87.

\bibitem{Hymer62} Hymer,~S. \& Pashigian,~P. (1962) {Firm size and rate of growth} {\it J.~of~Pol.~Econ.} 70:556-569.
% 
\bibitem{Mansfield62} Mansfield,~E. (1962) {Entry, Gibrat's law, innovation, and the growth of firms} {\it Amer.~Econ.~Rev.} 52:1023-1051.
% 
\bibitem{SimonD64} Simon,~H.~A. (1964) {Comment: Firm size and rate of growth} {\it J.~of~Pol.~Econ.} 72:81-82.
% 
\bibitem{HymerD64}  Hymer,~S. \& Pashigian,~P. (1964) {Firm size and rate of growth: Reply} {\it J.~of~Pol.~Econ.} 72:83-84.

\bibitem{Stanley96} Stanley~M~H~R, Amaral~L~A~N, Buldyrev~S~V, Havlin~S, Leschhorn~H, Maass~P, Salinger~M~A, 
Stanley~H~E (1996) {Scaling behavior in the growth of companies}. {\it Nature} 379:804-806.

\bibitem{Bottazzi01} Bottazzi~G, Dosi~G, Lippi~M, Pammolli~F \& Riccaboni~M (2001) {Innovation and corporate growth in the evolution of the drug industry}. \textit{Int J Ind Org} 19:1161-1187.

\bibitem{Sutton02} Sutton~J (2002) {The variance of firm growth rates: the \textquoteleft scaling\textquoteright~puzzle}. {\it Physica A} 312:577--590.

\bibitem{DeFabritiis03} De Fabritiis~G~D, Pammolli~F, Riccaboni~M (2003) {On size and growth of business firms}. {\it Physica A} 324:38--44.

\bibitem{Sergey_II} Buldyrev~S~V, Amaral~L~A~N, Havlin~S,
Leschhorn~H, Maass~P, Salinger~M~A, Stanley~H~E, Stanley~M~H~R (1997) {Scaling behavior in economics: II. modeling of company growth}. {\it J~Phys~I France} 7:635--650.

\bibitem{Amaral97} Amaral~L~A~N, Buldyrev~S~V, Havlin~S,
Leschhorn~H, Maass~P, Salinger~M~A, Stanley~H~E, Stanley~M~H~R (1997) {Scaling behavior in economics: I. empirical results for company growth}. {\it J Phys I France} 7:621--633.

\bibitem{Aoki07} Aoki~M, Yoshikawa~H (2007) {\it Reconstructing macroeconomics: a perspective from statistical physics and combinatorial stochastic processes} (Cambridge University Press, Cambridge, MA.).

\bibitem{Axtell06} Axtell~R (2006) {\it Firm sizes: facts, formulae and fantasies} (CSED Working Paper 44).

\bibitem{Klepper06} Klepper~S, Thompson~P (2006) {Submarkets and the evolution of market structure} {\it RAND J~Econ} 37:861--886.

\bibitem{Gabaix99} Gabaix~X (1999) {Zipf's law for cities: an explanation}. {\it Quar J Econ} 114:739--767.

\bibitem{Sutton07} Armstrong~M, Porter~R~H (2007) {\it Handbook of industrial organization, Vol. III} (North Holland, Amsterdam).

\bibitem{Fu_PNAS} Fu~D, Pammolli~F, Buldyrev~S~V, Riccaboni~M, Matia~K, Yamasaki~K, Stanley~H~E (2005) {The growth of business firms: theoretical framework and empirical evidence}. {\it Proc Natl Acad Sci USA} 102:18801.

\bibitem{Growiec07} Buldyrev~S~V, Growiec~G, Pammolli~F, Riccaboni~M, Stanley~H~E (2008) {The growth of business firms: facts and theory}. {\it J~Eu~Econ~Ass} 5:574-584.

\bibitem{Growiec08} Growiec~G, Pammolli~F, Riccaboni~M, Stanley~H~E (2008) {On the size distribution of business firms}. {\it Econ~Lett} 98:207-212.

\bibitem{Buldyrev07} Buldyrev~S~V, Pammolli~F, Riccaboni~M, Yamasaki~K, Fu~D, Matia~K, Stanley~H~E (2007) {A generalized preferential attachment model for business firms growth rates - II. Mathematical treatment}.
{\it Europ Phys J B} 57:131-138.

\bibitem{Pammolli07} Pammolli~F, Fu~D, Buldyrev~S~V, Riccaboni~M, Matia~K, Yamasaki~K, Stanley~H~E (2007)
{A generalized preferential attachment model for business firms growth rates - I. Empirical evidence} {\it Europ Phys J B} 57:127-130.

\bibitem{Amaral98} Amaral~L~A~N, Buldyrev~S~V, Havlin~S, Salinger~M~A, Stanley~H~E (1998) {Power law scaling for a system of interacting units with complex internal structure}. {\it Phys Rev Lett} 80:1385--1388.

\bibitem{Takayasu98} Takayasu~H, Okuyama~K (1998) {Country dependence on company size distributions and a numerical model based on competition and cooperation}. {\it Fractals\/} 6:67--79.

\bibitem{Canning98} Canning~D, Amaral~L~A~N, Lee~Y, Meyer~M, Stanley~H~E (1998) {Scaling the volatility of GDP growth rates}. {\it Econ~Lett} 60:335--341.

\bibitem{Buldyrev03} Buldyrev~S~V, Dokholyan~N~V, Erramilli~S, Hong~M, Kim~J~Y, Malescio~G, Stanley~H~E (2003) {Hierarchy in social organization}. {\it Physica A} 330:653--659.

\bibitem{Kazuko} Yamasaki~K, Matia~K, Buldyrev~S~V, Fu~D, Pammolli~F, Riccaboni~M, Stanley~H~E (2006) {Preferential attachment and growth dynamics in complex systems}. {\it Phys Rev E} 74:035103.

% \bibitem{Slimane01} Slimane,~S. (2001) {\it IEEE~Trans.~Comm.}
% {\bf 49}, 975-978.




\end{thebibliography}
\end{document}